\documentclass[times,twocolumn,final]{elsarticle}
\usepackage{medima}
\usepackage{framed,multirow}

\usepackage{xcolor}

\usepackage{amssymb}
\usepackage{latexsym}

\usepackage{url}
\usepackage{xcolor}

\usepackage{hyperref}
\usepackage{booktabs}
\definecolor{newcolor}{rgb}{.8,.349,.1}

\journal{preprint version}

\begin{document}

\verso{Zhixiang Zhang \textit{et~al.}}

\begin{frontmatter}

\title{Reconstructing high-order sequence features of dynamic functional connectivity networks based on diversified covert attention patterns for Alzheimer's disease classification}%

\author[1]{Zhixiang \snm{Zhang}\fnref{fn1}}

\author[1]{Biao \snm{Jie}\corref{cor1}}
\cortext[cor1]{Corresponding author: 
	E-mail: jbiao@ahnu.edu.cn;  }
\author[1]{Zhengdong \snm{Wang}}
\author[1]{Jie \snm{Zhou}}
\author[1]{Yang \snm{Yang}}
\author[1]{Wen \snm{Li}}
\author[1]{Zhaoxiang \snm{Wu}}

\address[1]{School of Computer and Information, Anhui Normal University, Anhui 241003, China}

\received{}
\finalform{}
\accepted{}
\availableonline{}
\communicated{}

\begin{abstract}
%%%
Recent studies have applied deep learning methods such as convolutional recurrent neural networks (CRNs) and Transformers to brain disease classification based on dynamic functional connectivity networks (dFCNs), such as Alzheimer's disease (AD), achieving better performance than traditional machine learning methods. However, in CRNs, the continuous convolution operations used to obtain high-order aggregation features may overlook the non-linear correlation between different brain regions due to the essence of convolution being the linear weighted sum of local elements. Inspired by modern neuroscience on the research of covert attention in the nervous system, we introduce the self-attention mechanism, a core module of Transformers, to model diversified covert attention patterns and apply these patterns to reconstruct high-order sequence features of dFCNs in order to learn complex dynamic changes in brain information flow. Therefore, we propose a novel CRN method based on diversified covert attention patterns, DCA-CRN, which combines the advantages of CRNs in capturing local spatio-temporal features and sequence change patterns, as well as Transformers in learning global and high-order correlation features. Experimental results on the ADNI and ADHD-200 datasets demonstrate the prediction performance and generalization ability of our proposed method.
%%%
\end{abstract}


\end{frontmatter}

%\linenumbers

%% main text
\section{Introduction}
Alzheimer’s disease (AD) is a major global health issue as an incurable neurodegenerative disease, and once diagnosed, the patient’s life expectancy can be shortened by 3 to 14 years. According to the World Health Organization (WHO), more than 55 million people worldwide currently have dementia, with AD being the most common form of dementia, contributing to 60-70\% of cases \citep{nandi2022global}. Studies into AD are, therefore, of great significance. Advances in our understanding of the genetic and pathological mechanisms underlying the disease, as well as the development of new diagnostic tools and treatments, have the potential to improve the lives of millions of people affected by AD and other forms of dementia \citep{gonzalez2023brain}. In AD, functional magnetic resonance imaging (fMRI) serves as a potential imaging biomarker that can preliminarily determine whether brain function is abnormal before structural changes occur in the brain \citep{david2022cognitive}, which is of great significance for the detection of early mild cognitive impairment (eMCI).

Numerous studies have demonstrated the potential of deep learning methods to become a new means of detecting AD in the future by using resting-state functional magnetic resonance imaging (rs-fMRI) data to construct dFCNs \citep{pan2023deep,khojaste2022deep}. However, these studies face various challenges and difficulties ranging from data collection to preprocessing. Despite the requirement for subjects to remain still and lying flat during rs-fMRI acquisition, it is impossible to control the diversity of spontaneous brain activity in subjects, including factors such as equipment noise and spontaneous thinking. After data collection, brain activity time series that has undergone strict preprocessing work such as registration, smoothing, removal of linear drift, and filtering still suffers from severe noise interference. In addition, due to hospitals' protection of patient privacy, studies also face the severe problem of data scarcity and uneven distribution of subject types \citep{forman2021targeting,van2019cognitive}. 
Therefore, the latest deep learning methods are focused to solve the problem of achieving better classification performance and generalization ability on limited clinical data from subjects as much as possible.

Inspired by modern neuroscience on the research of covert attention in the nervous system \citep{li2021different}, we propose that modeling the diversity of covert attention patterns is important for constructing high-order sequence features of brain information flow. Covert attention refers to the process by which the brain spontaneously selects and processes internal information in the absence of external stimuli or task demands \citep{petersen2012attention}, as well as the ability to control and change the focus of our attention \citep{culham2001attention,posner1990attention}. In other words, the diversity of covert attention patterns is one of the essential factors for complex dynamic changes in the brain information flow.
In real-world scenarios, even when subjects are physically still and lying flat in the acquisition device, the brain can still produce continuous switching of subjective consciousness under the influence of different covert attention patterns, such as unconscious active thinking and wandering \citep{aisen2022early,kamagata2022association}. The flow of information between different brain regions is constantly changing, like water flowing through winding caves, presenting complex dynamic patterns without external interference. However, it is not easy to capture these dynamic patterns through physical means; instead, it is more feasible to construct and study these dynamic patterns through mathematical algorithms.

Some recent studies have used graph neural network (GNN) models to construct dynamic change patterns of brain information flow to predict the classification of AD \citep{bi2023community,mishra2022graph,zhu2022interpretable}. These studies typically use brain regions as central nodes, with functional connections between brain regions as edges between central and adjacent nodes, to construct graphs and learn more complex high-order sequence features of dFCNs through graph convolution\citep{zhu2022interpretable}. However, there are also several potential drawbacks to the application of GNNs in the classification of AD: firstly, from a model design perspective, graph neural networks require a large amount of computational resources and time for feature construction and training on graph-structured data, especially when the scale and complexity of the graph are large \citep{bhatti2023deep,bessadok2022graph}. Secondly, from a model principle perspective, GNNs are usually based on some assumptions and approximations, such as fixed graph topology, homogeneous nodes and edges, symmetric adjacency matrix, etc. \citep{phan2023aspect}, which may not be applicable to the scenario of AD.

Unlike GNNs, convolutional neural networks (CNNs) can better avoid algorithmic complexity problems and graph structure assumption approximation due to their convolutional computation characteristics. In addition, CNNs have the specialities of locality and translation invariance \citep{kshatri2023convolutional,gu2018recent}, better capturing local dynamic change patterns of brain information flow along the spatio-temporal dimensions. Combined with the modeling of high-order sequence features of dFCNs using recurrent neural networks, convolutional recurrent neural networks (CRNs) persistently maintain advanced performance in the application of classification for AD \citep{lei2022longitudinal,jie2020designing}. CRN-based methods typically perform continuous convolution operations on the input dFCNs to obtain high-order brain network aggregation features, extract features from each sliding window using sequence segmentation, and combine recurrent neural network layers and fully connected layers for classification. Althgough, CRNs might neglect to consider the potential effect of the diversity of covert attention patterns on brain information flow, and the adequacy of this consideration is also veryfied in the experimental analysis of our study.

Therefore, we propose a novel CRN framework for AD classification based on diversified covert attention patterns (DCA-CRN). As one of the key kernels of the Transformer \citep{zhang2023applications}, the self-attention mechanism is used in our study to construct covert attention patterns in the complex dynamic changes of brain information flow. DCA-CRN retains the performance advantages brought by the structural design of the CRN and aggregates the high-order sequence features of dFCNs constructed by the sliding window method. It achieves commendable classification performance on both ADNI and ADHD-200 datasets.

This paper has three main contributions. First, we propose using the self-attention mechanism to construct diversified covert attention patterns in brain information flow and apply these patterns to reconstruct the high-order sequence features of dFCNs. Second, we further validate the proposed method on the multi-class classification task of AD and conduct a comparative analysis of the differences in disease-specific brain regions between the CNN, CRN, Transformer, and DCA-CRN. Lastly, we preliminarily inspect the similarities and differences of covert attention patterns of subjects at different stages (NC et al.) by visualizing the attention score heatmaps.

\section{Related work}
\subsection{Funcational connectivity network (FCN) construction}
Functional magnetic resonance imaging (fMRI) offers the capability to detect brain abnormalities beyond the scope of other imaging techniques, particularly when subtle changes occur without significant structural alterations \citep{matthews2016clinical}. This advanced imaging modality enables the exploration of functional brain anatomy, assessment of brain connectivity, and evaluation of the impact of conditions like stroke or other diseases \citep{matthews2006applications}. Notably, alterations in functional connectivity networks (FCNs) have emerged as potential biomarkers for categorizing and predicting neurological disorders. In the domain of brain disorder diagnosis, the recent focus has been on investigating resting-state fMRI (rs-fMRI) derived FCNs, as demonstrated by studies on disorders such as AD \citep{qiu2022multimodal,kam2019deep}. These investigations often categorize FCN construction into two primary types: static FCNs (sFCNs) and dynamic FCNs (dFCNs), leading to a classification of research efforts aligned with these construction methodologies.

Currently, for sFCN, studies have been dedicated to the longitudinal study of brain disease diagnosis, using multi-time points rs-fMRI data. For example, \citet{yang2019fused} proposed a fusion sparse network (FSN) method to extract longitudinal features for detecting MCI phases. \citet{huang2020} proposed to add both Pearson correlation correction (PCC) and modular structure to the sparse low-rank brain network (SLR), obtaining the PCC-related SLR brain network features for Autism Spectrum Disorder (ASD) diagnosis. However, they only focused on sFCNs, thus neglecting the dynamic patterns of brain information flow. Another potential problem with the longitudinal study is the need for a years-long accumulation of clinical data, which may be detrimental to early diagnosis of brain diseases. As a result, the dynamic patterns of dFCNs may provide better prospects for clinical applications in the early diagnosis of AD and personalized medicine while facilitating a better understanding of the pathological basis.

To investigate complex dynamic changes of dFCNs over time, existing studies often use analysis of sliding window correlations among brain regional activities to estimate correlations of brain activity across multiple, possibly overlapping, time-series segments. For example, \citet{chen2017extraction} extract the root-mean-square features of dFCNs for automatic diagnosis of mild cognitive impairment (MCI) via ensemble support vector machine (SVM). \citet{zhao2021diagnosis} proposed a new method that integrates multiple functional magnetic resonance imaging (fMRI) views to create more accurate functional connectivity networks. This network construction method helps capture the complex functional changes within the brains of ASD patients, thereby improving the accuracy of diagnosis. The research results show that this method has potential significant advantages in the diagnosis of ASD, providing new ideas for the development of more accurate diagnostic tools and treatment methods in the future. Since dFCNs retain the sequential nature of the original fMRI data and may reflect more complex dynamic pattern information, many cutting-edge studies are predicting the classification of AD via deep learning methods \citep{pan2023deep,huang2023sd,jie2020designing}.

\subsection{Connectivity-network-based learning methods for AD classification}

In recent years, FCN-based learning methods for brain disease classification have been continuously upgraded in the in-depth application of machine learning and deep learning technologies. According to existing studies, we can divide their main development history into four stages: the first stage is the exploration stage combining feature selection with machine learning classifiers; the second stage is the early stage of deep learning with CNNs as the baseline method; the third stage is the mid-stage of deep learning dominated by CRNs and GNNs; and the fourth stage is the new era of deep learning with Transformers as the new benchmark method.

During the exploration phase, different feature selection methods (e.g., Recursive Feature Elimination (RFE) \citep{wee2011enriched}, t-test) were used to filter meaningful features from the constructed sFCNs and combined with classical machine learning methods (e.g., Support Vector Machine (SVM) \citep{noble2006support}) to improve the classification performance of brain diseases. These methods include extracting local network features for brain disease classification, adopting more advanced methods to learn global network features, and integrating local and global network features. For example, \citet{chen2011classification} constructed sFCNs based on rs-fMRI data, then extracted local network features such as node degree and clustering coefficient and combined them with large-scale network analysis to classify different stages of AD. In addition, \citet{uddin2013salience} first extracted local network features of node strength and node efficiency of significant networks and further used SVM to classify and predict the severity of symptoms of ASD in children. However, these methods only considered local network features and ignored the global topological properties of the network. Therefore, \citet{jie2014topological} proposed a topological graph kernel method based on global network features to measure the similarity between multiple thresholded dFCNs in the classification of Mild Cognitive Impairment (MCI). Some studies also aggregate local and global topological features to improve the classification performance of brain diseases \citep{chen2017hierarchical}. All studies suggest that while considering the local dynamic changes of brain information flow along the time dimension, its global topological features along the spatial dimension are also significant.

In the early stage of deep learning, CNNs were applied to classify brain diseases and significantly improved performance compared with traditional machine learning methods\citet{sri2022detection}. CNNs have the characteristics of locality and translation invariance, making them better at capturing local dynamic changes in brain information flow along the spatio-temporal dimensions. For example, \citet{kawahara2017brainnetcnn} proposed a novel framework based on CNNs for predicting neurodevelopment using brain network connectivity data. Considering that the interactions between brain regions may have different degrees of importance at different time points, \citet{jie2020designing} effectively constructed dFCNs from rs-fMRI using a weighted correlation kernel model and achieved superior classification performance on the ADNI and ADHD-200 datasets. In addition, \citet{huang2023sd} proposed a static-dynamic convolutional neural network to fully utilize the advantages of sFCNs and dFCNs, achieving excellent performance on two real epilepsy and schizophrenia datasets. Although many CNN-based classification methods use dFCNs as input, they ignore the consideration of global dynamic changes in dFCNs along the temporal dimension.

During the mid-stage of deep learning, CRNs and GNNs were applied to classify brain diseases, and CRNs became the mainstream method with stable performance advantages\citet{shoeibi2022diagnosis}. By modeling the high-order sequence features of dFCNs with recursive neural networks, CRNs achieved excellent performance in processing spatio-temporal sequence information. For instance, \citet{yan2019discriminating} developed a multi-scale model based on Recurrent Neural Networks (RNNs) that directly classified 558 schizophrenia patients and 542 healthy controls using the time series of fMRI independent components (ICs). \citet{zhao2022attention} proposed an attention-based hybrid deep learning framework that combined the brain connectivity and temporal consistency and dynamics of brain activity from fMRI data to distinguish between schizophrenia patients and healthy controls. Unlike CRNs, GNNs were also used for brain disease classification based on connectome data due to their ability to effectively handle unstructured and unordered brain data. \citet{zhang2022classification} proposed a Local-to-Global Graph Neural Network (LG-GNN) to learn the feature embeddings of local brain regions using the generated embeddings and non-imaging information to learn the relationships between subjects, achieving classification of brain diseases such as ASD and AD. Despite GNNs’ ability to better construct the local and global spatial topological patterns of dFCNs from a graph theory perspective, with some proposed methods achieving a certain degree of interpretability, GNNs still face issues such as the need for assumptions and approximations, as well as high training time overhead and computational complexity.

In the new era of deep learning, the Transformer, a powerful and flexible learning network framework, has been applied to classify brain diseases. Transformers can effectively use the self-attention mechanism to model the dynamic correlations of dFCNs, capturing the complex dynamic changes in brain information flow. Specifically, \citet{zhang2022diffusion} proposed an improved method based on the Transformer, using dFCNs constructed from resting-state functional magnetic resonance imaging (rs-fMRI) data to model, analyze and classify brain diseases such as Attention Deficit Hyperactivity Disorder (ADHD) and AD. However, current studies based on Transformers have only been validated for binary classification tasks in brain diseases. There is a lack of comparative analysis between CNNs, CRNs, and Transformers in brain-disease-specific region differences and further performance validation of multi-class tasks. In addition, to address the large amount of training data required by the Transformer architecture, we propose simplifying our method's original self-attention mechanism in this paper.

We conduct relevant experiments and analyses for the AD classification using the ADNI dataset to validate our proposed method. As a supplement, we further validate the generality of DCA-CRN for classifying brain diseases on the ADHD-200 dataset. 

\begin{table}[!ht]
	\scriptsize
	\renewcommand{\arraystretch}{1.2}
	\caption{Characteristics of the studied subjects (Mean $\pm$ Standard Deviation). MMSE: Mini-Mental State Examination.}
	\begin{tabular*}{0.48\textwidth}{@{\extracolsep{\fill}}l ccccc}
		\toprule
		~Dataset & Group & \# Subjects & Scan & Male/Female & Age \\
		\midrule
		\multirow{4}{*}{ADNI} & AD   & 31 & 99  & 16/15 & $74.7 \pm 7.4$ \\
		& lMCI & 45 & 145 & 27/18 & $72.3 \pm 8.1$ \\
		& eMCI & 50 & 165 & 20/30 & $72.4 \pm 7.1$ \\
		& NC   & 48 & 154 & 20/28 & $76.0 \pm 6.8$ \\
		\midrule
		\multirow{2}{*}{ADHD-200} & ADHD & 118 & 436 & 25/93 & $11.2 \pm 2.7$ \\
		& NC   & 98  & 356 & 51/47 & $12.2 \pm 2.1$ \\
		\bottomrule
	\end{tabular*}
	\label{tab1_subjects}
\end{table}

\section{Method}
\subsection{Subjects and fMRI data preprocessing}

We use rs-fMRI data for $174$ subjects, including $48$ normal controls (NCs), $50$ eMCI, $45$ late mild cognitive impairment (lMCI), and $31$ AD, which come from the ADNI database. Of these, $154$, $165$, $145$, and $99$ scans obtained from each of the nine different periods are for AD, lMCI, eMCI, and NCs, respectively. Nevertheless, data of the eMCI and AD groups are unfilled from month $48$ to month $84$. The scan parameters for data are as follows: the echo time (TE) is $30\,ms$, $2.2$-$3.1\,s$ for the repetition time (TR), $2.29$-$3.31\,mm$ for the in-plane image resolution and $3.31\,mm$ for slice thickness. Following the previous study approach \citep{jie2018integration}, we adopt a standard processing flow with the FSL FEAT software package~\footnote{https://fsl.fmrib.ox.ac.uk/fsl/fslwiki/FEAT}: (1) Remove the first three volumes of sampled data. (2) Do slice timing. (3) Correct head motion. (4) Do bandpass filtering. 
(5) Regress white matter, CSF, and motion parameter covariates.  We divide the subject's brain space of rs-fMRI scans into 116 brain regions by the specified AAL template and the non-parametric registration method. At last, we treat the average time series calculated from BOLD signals in each brain region as the input.

The demographic and clinical information of these subjects is summarized in Table~\ref{tab1_subjects}, indicating that our study's amount and proportion of subjects for the AD classification are well balanced. In addition, AD subjects are mainly aged between 64 and 83 using the ADNI dataset, which suggests that our findings hold more pertinence and significance for an older population at high risk of AD \citep{scheltens2021alzheimer}.

To better validate the applicability of our method to the classification of different brain diseases, we also conducted performance validation on the ADHD-200 dataset, as shown in Section~\ref{sec:adhd}. Also, in Table~\ref{tab1_subjects}, the ages of subjects with ADHD are mainly concentrated in the range of 8 to 15 years old, which is consistent with existing clinical statistical research results \citep{hinshaw2018attention,faraone2003worldwide}. 

\subsection{Reconstructing high-order sequence features based on diversified covert attention patterns}

Before making use of the self-attention mechanism to construct covert attention patterns, we scrutinized the feature extraction phase in previous CNN or CRN-based methods for AD classification. They usually perform successive convolution operations on the input dFCNs to obtain high-order brain network aggregation features. However, if you further observe the nature of convolution, you will find that convolution is a linear feature aggregation method that ignores the nonlinearity of information superposition because multiple brain regions interact with each other \citep{perovnik2023functional,rabinovich2012information}. Additionally, according to recent studies, a vital characteristic of the self-attention mechanism is its data specificity rather than long-range dependency \citep{merrill2023parallelism,paul2022vision}, which means that it is capable of uncovering intrinsic correlations between data features. This characteristic provides a neoteric way for modeling complex dynamic changes in brain information flow. Therefore, we employ the self-attention mechanism to construct diversified covert attention patterns and apply these patterns to reconstruct the high-order sequence features of dFCNs.

Given the high-order sequence features $\mathbf{I\epsilon R}^{\mathbf{m}\times \mathbf{n}\times \mathbf{c}}$ of dFCNs, where $\mathbf{m}$ refers to the total length of time windows, $\mathbf{n}$ represents the number of brain regions, and $\mathbf{c}$ represents the number of high-order sequence feature channels. First, we create trainable linear weight vectors $\mathbf{W}_{\mathbf{Q}},\mathbf{W}_{\mathbf{K}},\mathbf{W}_{\mathbf{V}}\mathbf{\epsilon R}^{\mathbf{n}\times \mathbf{m}\times \mathbf{c}}$. Then, $\mathbf{I}$ is processed along the $\mathbf{c}$ dimension as follows: we obtain the query vector $\mathbf{Q}^{\mathbf{\lambda}}=\mathbf{I}^{\mathbf{\lambda}}\mathbf{W}_{\mathbf{Q}}^{\mathbf{\lambda}}$, the key vector $\mathbf{K}^{\mathbf{\lambda}}=\mathbf{I}^{\mathbf{\lambda}}\mathbf{W}_{\mathbf{K}}^{\mathbf{\lambda}}$, and the value vector $\mathbf{V}^{\mathbf{\lambda}}=\mathbf{I}^{\mathbf{\lambda}}\mathbf{W}_{\mathbf{V}}^{\mathbf{\lambda}}$, where $\mathbf{\lambda\epsilon }\left[ 1,\mathbf{c} \right]$. Next, we obtain the covert attention pattern vector of high-order sequence feature channel $\mathbf{\lambda}$ as $\mathbf{P}^{\mathbf{\lambda}}=\mathbf{Q}^{\mathbf{\lambda}}\left( \mathbf{K}^{\mathbf{\lambda}} \right) ^{\mathbf{T}}$. Then, the covert attention pattern vector $\mathbf{P}^{\mathbf{\lambda}}$ is divided by $\sqrt{\mathbf{d}_{\mathbf{k}}}$ ($\mathbf{d}_{\mathbf{k}}=\mathbf{n}$), and row normalization is performed to obtain the score vector $\mathbf{P}^{\mathbf{\lambda}}$ of the covert attention pattern $\mathbf{\lambda}$. Finally, the score vector $\mathbf{P}^{\mathbf{\lambda}}$ is multiplied by the value vector $\mathbf{V}^{\mathbf{\lambda}}$ to obtain the reconstructed high-order sequence features of dFCNs, $\mathbf{O}^{\mathbf{\lambda}}\mathbf{\epsilon R}^{\mathbf{m}\times \mathbf{n}}$. The process of constructing a covert attention pattern $\mathbf{\lambda}$ using the self-attention mechanism and further reconstructing features can be summarized as follows:
\begin{equation}
	\mathbf{O}^{\mathbf{\lambda }}=\mathbf{Soft}\max \left( \frac{\mathbf{Q}^{\mathbf{\lambda }}\left( \mathbf{K}^{\mathbf{\lambda }} \right) ^{\mathbf{T}}}{\sqrt{\mathbf{d}_{\mathbf{k}}}} \right) \mathbf{V}^{\mathbf{\lambda }}
	\label{equ1}
\end{equation}

\begin{figure*}[!htbp]
	\begin{center}
		\centering
		\includegraphics[width=1.0\textwidth]{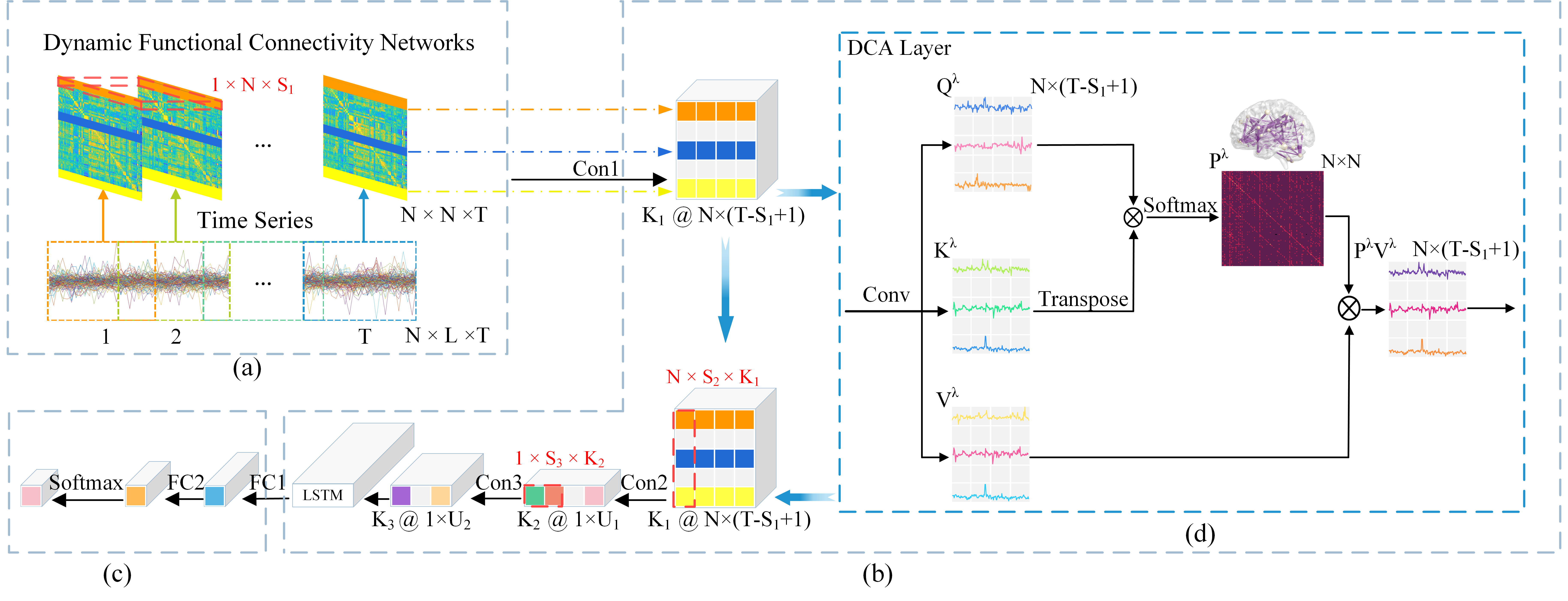}
		\caption{Illustration of the proposed DCA-CRN learning framework for high-order sequential feature extraction and classification with rs-fMRI data, including three parts: (a) dFCN construction, (b) spatial feature and temporal feature extraction, and (c) classification. Of these, (d) represents the complete course of the DCA layer.}
		\label{fig_frame}
	\end{center}
\end{figure*}

Specifically, we elaborate on the interpretability of reconstructing the high-order sequence features by modeling covert attention patterns. Taking the query vector $\mathbf{Q}^{\mathbf{\lambda }}\mathbf{\epsilon R}^{\mathbf{N}\times \left( \mathbf{T}-\mathbf{S}_1+\mathbf{1} \right)}$ as an example, $\mathbf{Q}_{\mathbf{i}\cdot}^{\mathbf{\lambda }}=\left[ \mathbf{Q}_{\mathbf{ij}}^{\mathbf{\lambda }} \right] _{\mathbf{j}=\mathbf{1}}^{\mathbf{T}-\mathbf{S}_{\mathbf{1}}+\mathbf{1}}$ represents the entire temporal features of the brain region corresponding to index $\mathbf{i}$, where $\mathbf{i\epsilon }\left[ \mathbf{1},\mathbf{N} \right] $. Similarly, $\mathbf{Q}_{\cdot \mathbf{j}}^{\mathbf{\lambda }}=\left[ \mathbf{Q}_{\mathbf{ij}}^{\mathbf{\lambda }} \right] _{\mathbf{i}=\mathbf{1}}^{\mathbf{N}}$ refers to spatial features of the whole brain regions within a single time window, where $\mathbf{j\epsilon }\left[ \mathbf{1},\mathbf{T}-\mathbf{S}_1+\mathbf{1} \right] $. We obtain $\mathbf{P}_{\mathbf{ij}}^{\mathbf{\lambda }}=\frac{\mathbf{Q}_{\mathbf{i}\cdot}^{\mathbf{\lambda }}\left( \mathbf{K}_{\cdot \mathbf{j}}^{\mathbf{\lambda }} \right) ^{\mathbf{T}}}{\sum_{\mathbf{j}=\mathbf{1}}^{\mathbf{T}-\mathbf{S}_1+\mathbf{1}}{\mathbf{Q}_{\mathbf{i}\cdot}^{\mathbf{\lambda }}}\left( \mathbf{K}_{\cdot \mathbf{j}}^{\mathbf{\lambda }} \right) ^{\mathbf{T}}}$, where $\mathbf{P}_{\mathbf{ij}}^{\mathbf{\lambda }}$ denotes the scores of the covert attention pattern $\mathbf{\lambda}$ corresponding to the spatio-temporal features of brain regions $\mathbf{i}$ and $\mathbf{j}$. Thus, the updated features of the $\mathbf{i}_\mathbf{th}$ brain region are $\mathbf{P}^{\mathbf{\lambda }}\mathbf{V}_{\mathbf{i}\cdot}^{\mathbf{\lambda }}=\left[ \sum_{\mathbf{n}=\mathbf{1}}^{\mathbf{N}}{\mathbf{P}_{\mathbf{in}}^{\mathbf{\lambda }}}\mathbf{V}_{\mathbf{nj}}^{\mathbf{\lambda }} \right] _{\mathbf{j}=\mathbf{1}}^{\mathbf{T}-\mathbf{S}_1+\mathbf{1}}$, where $\mathbf{P}^{\mathbf{\lambda }}\mathbf{V}_{\mathbf{i}\cdot}^{\mathbf{\lambda }}$ is the reconstructed high-order sequence features corresponding to the covert attention pattern $\mathbf{\lambda}$. 

In order to make the high-order sequence feature reconstruction process based on diversified covert attention patterns more applicable to few-shot learning, we introduce the following two adjustments: On the one hand, we adopt a self-attention mechanism based on 1 × 1 convolution as a linear transformation method, reducing the number of trainable weight parameters from 3 × c × n × m to 3 × c × 1 × 1. On the other hand, we adopt a residual network structure, adding the input vector $\mathbf{I}$ and output vector $\mathbf{O}$, utilizing the reconstructed features reconstructed through diversified covert attention weights to update the high-order sequence features of the input dFCNs, as shown in Fig.~\ref{fig_frame}~(d).

\subsection{Proposed DCA-CRN learning framework}
As shown in Fig.~\ref{fig_frame}, our proposed DCA-CRN learning framework consists of three parts: (a) dFCN construction, (b) spatial feature and temporal feature extraction, and (c) classification. We will cover the details of these three parts in the following subsections.

\subsubsection{dFCN construction} 
As shown in Fig.~\ref{fig_frame}~(a), 
dFCNs based on continuous and overlapping time windows are constructed by using the average time series of the specified  $\mathbf{N}$ brain regions. For every subject, we first segment his or her all rs-fMRI time series into  $\mathbf{T}$ consecutive and overlapping time windows of constant  $\mathbf{L}$ length. Next, we construct the dFCNs: $\mathbf{F}^{\mathbf{t}}\mathbf{\epsilon R}^{\mathbf{N}\times \mathbf{N}}\left( \mathbf{t}=1,\cdot \cdot \cdot ,\mathbf{T} \right)  $ with Pearson's correlation coefficients among BOLD signals of paired brain regions at the t-th time window, as shown below:
\begin{equation}
	\mathbf{F}^{\mathbf{t}}\left( \mathbf{j},\mathbf{k} \right) =\frac{\mathbf{covr}\left( \mathbf{x}_{\mathbf{j}}^{\mathbf{t}},\mathbf{x}_{\mathbf{k}}^{\mathbf{t}} \right)}{\mathbf{\sigma }_{\mathbf{x}_{\mathbf{j}}^{\mathbf{t}}}\mathbf{\sigma }_{\mathbf{x}_{\mathbf{k}}^{\mathbf{t}}}}\,\,                       
	\label{equ2}
\end{equation}
where $covr$ denotes the covariance of two vectors, $\mathbf{\sigma }_{\mathbf{x}_{\mathbf{j}}^{\mathbf{t}}}$ and $\mathbf{\sigma }_{\mathbf{x}_{\mathbf{k}}^{\mathbf{t}}}$ represent the standard deviation corresponding to the vectors $\mathbf{x}_{\mathbf{j}}^{\mathbf{t}}$ and $\mathbf{x}_{\mathbf{k}}^{\mathbf{t}}$ respectively. The vector $\mathbf{x}_{\mathbf{j}}^{\mathbf{t}}$ refers to the BOLD signals of the $\mathbf{j}_{\mathbf{th}}$ brain region at the $\mathbf{t}_{\mathbf{th}}$ time window as well as $\mathbf{x}_{\mathbf{k}}^{\mathbf{t}}$.

According to Eq.~\ref{equ1}, $\mathbf{F}^{\mathbf{t}}$ is a correlation coefficient matrix for all brain regions at the t-th time window. Further, each row or column of the matrix $\mathbf{F}^{\mathbf{t}}$ represents the degree of correlations between the specific brain region and all brain regions at time window t. Thus, a group of dFCNs $\mathbf{F}=\left\{ \mathbf{F}^1,\mathbf{F}^2,\cdot \cdot \cdot ,\mathbf{F}^{\mathbf{T}} \right\} \mathbf{\epsilon R}^{\mathbf{T}\times \mathbf{N}\times \mathbf{N}}$ work on transmitting dynamic brain information of these subjects.
\subsubsection{Spatial feature and temporal feature extraction}
First, we set the kernel size of the first convolution layer $\mathbf{Con}1$ to $1\times \mathbf{N}\times \mathbf{S}_1$, where $1\times \mathbf{N}$ represents the size of the convolution kernel for the spatial dimension and $\mathbf{S}_1$ indicates the size for the temporal dimension and set the stride to $\left( \mathbf{N},\mathbf{S}_1 \right) $ along the spatial dimension and the temporal dimension, mainly to aggregate information from separate brain regions across $\mathbf{S}_1$ adjacent time windows. 

Then, we reconstruct and update high-order sequence features of dFCNs channel-wise based on diversified covert attention patterns, as shown in Fig.~\ref{fig_frame}~(d). The $\mathbf{DCA}$ layer operates by the following process: 

We perform linear transformations of the input $\mathbf{I}^{\mathbf{\lambda }}\mathbf{\epsilon R}^{\mathbf{N}\times \left( \mathbf{T}-\mathbf{S}_1+1 \right)}$ from the $\mathbf{\lambda}_\mathbf{th}$  channel through thrice independent $1\times 1$ convolutions, where $\mathbf{N}$ denotes the amount of brain regions and $\left( \mathbf{T}-\mathbf{S}_1+1 \right) $ refers to the total length of time windows. So far, we obtain the vectors $\mathbf{Q}^{\mathbf{\lambda }},\mathbf{K}^{\mathbf{\lambda }},\mathbf{V}^{\mathbf{\lambda }}\mathbf{\epsilon R}^{\mathbf{N}\times \left( \mathbf{T}-\mathbf{S}_1+1 \right)}$ , where $\mathbf{Q}^{\mathbf{\lambda }}$ is used as the query vector, $\mathbf{K}^{\mathbf{\lambda }}$ as the key vector and $\mathbf{V}^{\mathbf{\lambda }}$ as the value vector. 
Subsequently, the query vector $\mathbf{Q}^{\mathbf{\lambda }}$ is multiplied by the transpose of the key vector $\mathbf{K}^{\mathbf{\lambda }}$ to yield the score vector $\mathbf{P}^{\mathbf{\lambda }}\mathbf{\epsilon R}^{\mathbf{N}\times \mathbf{N}}$. The score vector $\mathbf{P}^{\mathbf{\lambda }}\mathbf{\epsilon R}^{\mathbf{N}\times \mathbf{N}}$ signifies the dynamic correlation degree among all brain regions, while concurrently representing the covert attention pattern $\mathbf{\lambda }$ corresponding to the high-order sequence features of dFCNs. 
Then, the score vector $\mathbf{P}^{\mathbf{\lambda }}$ is divided by $\mathbf{d}_{\mathbf{K}}=\mathbf{T}-\mathbf{S}_1+1$ and normalized by Softmax function, as in equation $\mathbf{P}^{\mathbf{\lambda }}=\mathbf{Soft}\max \left( \frac{\mathbf{Q}^{\mathbf{\lambda }}\left( \mathbf{K}^{\mathbf{\lambda }} \right) ^{\mathbf{T}}}{\sqrt{\mathbf{d}_{\mathbf{k}}}} \right)$ . Unlike Pearson's correlations which construct relevant degree for paired brain regions in a complex calculation (Eq.~\ref{equ2}), it is more efficient to get dot-product of $\mathbf{Q}^{\mathbf{\lambda }}$ and the transpose of $\mathbf{K}^{\mathbf{\lambda }}$ in much cheaper computation complexity (Eq.~\ref{equ1}). Moreover, contrary to traditional convolutions that linearly perform feature aggregation, the vectors $\mathbf{Q}^{\mathbf{\lambda }}$ and $\mathbf{K}^{\mathbf{\lambda }}$ are learnable so that the score vector $\mathbf{P}^{\mathbf{\lambda }}$ can be non-linearly updated as the input $\mathbf{I}^{\mathbf{\lambda }}$ changes, updating dynamic correlated information in the high order sequence features from different brain regions.

For the final step of the $\mathbf{DCA}$ layer, we introduce the residual block \citep{szegedy2017inception} to the output, aggregating both the input and reconstructed features, as expressed in equation $\mathbf{O}^{\mathbf{\lambda }}=\mathbf{P}^{\mathbf{\lambda }}\mathbf{V}^{\mathbf{\lambda }}+\mathbf{I}^{\mathbf{\lambda }}$, where $\mathbf{P}^{\mathbf{\lambda }}\mathbf{V}^{\mathbf{\lambda }}$ indicates the value vector $\mathbf{V}^{\mathbf{\lambda }}$ should be reconstructed with the score vector $\mathbf{P}^{\mathbf{\lambda }}$, and $\mathbf{O}^{\mathbf{\lambda }}$ stands for the output. 

After features from all channels have been reconstructed, layer normalization (LN) serves as the activation function of the $\mathbf{DCA}$ layer. 	

Then, convolution layers $\mathbf{Con}2$ and $\mathbf{Con}3$ are utilized to further aggregate high-order features from different brain regions along spatial and temporal dimensions. The kernel size of the layer $\mathbf{Con}2$ equals to $\mathbf{N}\times \mathbf{S}_2\times \mathbf{K}_1$, and the kernel size of the layer $\mathbf{Con}3$ is $1\times \mathbf{S}_3\times \mathbf{K}_2$. We accordingly set the stride of the layer $\mathbf{Con}2$ and $\mathbf{Con}3$ to ($1$, $1$) and ($1$, $2$), respectively. Each convolutional layer ($\mathbf{Con}1$, $\mathbf{Con}2$, $\mathbf{Con}3$) is followed by batch normalization (BN), rectified linear unit (ReLU) activation, and $0.25$ dropout. 

At the end of the spatial feature and temporal feature extraction stage, to capture the sequence change patterns and dig deeper into the different contributions among time series of dFCNs, we set LSTM (containing 48 neurons) block followed by RELU activation and 0.5 dropout.  

\subsubsection{Classification}
As shown in Fig.~\ref{fig_frame}~(c), we employ two fully connected layers and a softmax layer for prediction with the output of LSTM as input. In particular, we set up $32$ neurons and $16$ neurons for the first fully connected layer and the second fully connected layer, respectively. There are $2$ and $4$ neurons in the last fully connected layer for binary and multi-class classification separately. It is worth noting that we adopt L2 regularization in the last fully connected layer for better parameter optimization.

\section{Experiments and results}
\subsection{Experimental setup}
The study uses a 5-fold cross-validation (CV) strategy to ensure that the training set and test set do not overlap. We perform both binary and multi-class classification experiments, including (1) eMCI vs. NC classification, (2) AD vs. NC classification, and (3) AD vs. lMCI vs. eMCI vs. NC classification. For each classification task, we expressly partition all subjects into five subsets of the same size. Each subset is selected as the test set, while the remaining four subsets are joined together to construct the training set. Further, we select one-fifth of the training set as validation data to find the best empirical parameters for the optimal model. It is worth pointing out that each scan of each subject is treated as an independent sample to enhance the model's generalization ability, but all scans of the same subject have the same class label. 

For our proposed network shown in Fig.~\ref{fig_frame}, we empirically set the parameters as follows: $\mathbf{N}$ = $116$, $\mathbf{T}$= $34$, $\mathbf{L}$= $70$, $\mathbf{S}_1$ = $2$, $\mathbf{S}_2$ = $1$, $\mathbf{S}_3$ = $8$, $\mathbf{K}_1$ = $5$, $\mathbf{K}_2$ = $16$, $\mathbf{K}_3$ = $32$, $\mathbf{U}_1$ = $33$, $\mathbf{U}_2$ = $13$. The Adam optimizer with recommended parameters is used for training, and the number of epochs and batch size are empirically set as $200$ and $16$, respectively.

We compare our method with the following methods. (1) \textbf{Baseline}: The sFCNs are constructed by computing the Pearson correlation coefficient among the time series of each ROI at the start. Next, the connectivity strengths of the sFCNs are set to the features. After feature selection with the t-test method (i.e., p-value $<$ 0.05), a linear SVM with default parameters is used for classification. (2) \textbf{SVM}: Contrary to the Baseline, local clustering coefficients of the sFCNs from each subject are extracted as features. Then, the t-test method and a linear SVM with default parameters are also used for feature selection and classification, respectively. (3) \textbf{DFCN-mean}: The dFCNs are first constructed for each subject. Next, the temporal and spatial mean features of dFCNs are extracted. The manifold regularized multi-task feature learning (M2TFL) and multi-kernel SVM are used for feature selection and classification, respectively. (4) \textbf{CNN}: The dFCNs are constructed for each subject as the input. Then, consecutive convolution layers and an average pooling layer are used to extract features. Finally, fully connected layers (FCs) and a softmax layer are used for classification. (5) \textbf{CRN}: As a variant of the CNN, this method has a similar network architecture but replaces the LSTM layer with an average pooling layer considering the temporal dynamics along with time steps. (6) \textbf{Transformer}: First, dFCNs are fed to two successive convolutional layers for patch embedding. Transformer's encoder consists of MSA, a residual block, and FCs.Then, the cascaded Transformer blocks are set to extract high-order spatio-temporal features. After that, an average pooling layer is used to extract features. Finally, the classification part of the Transformer keeps the same settings as the CRN.   

\begin{table*}[!tbp]	
	\scriptsize
	\renewcommand{\arraystretch}{1.2}
	\caption{Performance of seven methods in two binary classification tasks, i.e., eMCI vs. NC and AD vs. NC classifications. ACC=Accuracy.} 
	\begin{center}
		\begin{tabular*}{1.0\textwidth}{@{\extracolsep{\fill}}l ccc cccc}
			\toprule
			\multirow{2}{*}{~Method}&\multicolumn{3}{c}{eMCI vs. NC (\%)} &&\multicolumn{3}{c}{AD vs. NC (\%) } \\
			\cline{2-4} \cline{6-8} 
			& ACC & SPE  & SEN  && ACC & SPE & SEN\\
			\hline
			~Baseline & $57.1$ & $48.1$ & $65.6$ && $73.3$ & $77.8$ & $66.7$\\
			
			~SVM & $63.6$ & $50.0$ & $75.0$ && $75.0$ & $80.0$ & $66.7$\\
			
			~DFCN-mean & $67.7$ & $47.3$ & $84.7$ && $76.4$ &\textbf{$100.0$} & $33.3$\\
			
			~CNN  & $80.8$ & $88.4$ & $78.6$ && $87.8$ & $95.0$ & $88.1$\\
			
			~CRN  & $84.2$ & $84.7$ & $88.3$ && $92.8$ & $95.0$ & $93.8$\\
			
			~Transformer & $86.3$ & \textbf{$90.5$} & $81.3$ && $92.8$ & $80.0$ & \textbf{$100.0$}\\
			
			~DCA-CRN(OURS)   & \textbf{$87.8$} & $89.7$ & \textbf{$88.7$} && \textbf{$97.5$}&\textbf{$100.0$} & $96.7$\\	
			\bottomrule

		\end{tabular*}
	\end{center}
	\label{tab2_acc}
\end{table*} 

\begin{table*}[!tbp]	
	\setlength{\belowcaptionskip}{0pt}
	\centering\renewcommand{\arraystretch}{1.2}
	\caption{Performance of seven methods in the multi-class classification task, i.e.,  AD vs. lMCI vs. eMCI vs. NC classification. ACC=Accuracy.} 
	\scriptsize
	\begin{center}
		\begin{tabular*}{1.0\textwidth}{@{\extracolsep{\fill}}l ccc cc}
			\toprule
			\multirow{2}{*}{~Method}&\multicolumn{5}{c}{AD vs. lMCI vs. eMCI vs. NC (\%)}  \\
			\cline{2-6} 
			& ACC & ACC$_{NC}$ & ACC$_{eMCI}$ & ACC$_{lMCI}$ & ACC$_{AD}$\\
			\hline
			~Baseline & $30.6$ & $20.0$ & $38.9$ & $30.0$ & $33.3$\\
			
			~SVM & $35.0$ & $22.0$ & $69.5$ & $21.0$ & $6.7$\\
			
			~DFCN-mean & $44.0$ & $36.0$ & \textbf{$87.6$} & $22.0$ & $0.0$\\
			
			~CNN  & $62.2$ & $21.4$ & $74.8$ & $52.0$ & $60.0$\\
			
			~CRN  & $67.4$ & $52.7$ & $74.8$ & $64.0$ & \textbf{$66.7$}\\
			
			~Transformer & $65.7$ & $55.3$ & $66.2$ & \textbf{$77.0$} & \textbf{$66.7$}\\
			
			~DCA-CRN(OURS)   &\textbf{$69.6$}&\textbf{$64.7$}& $78.1$ & $65.0$ &\textbf{$66.7$}\\	
			\bottomrule		
			\label{tab3_acc}
		\end{tabular*}
	\end{center}
\end{table*}

\subsection{Classification performance}

On the one hand, Table~\ref{tab2_acc} and Table~\ref{tab3_acc} depict comparison results of all methods for two binary and multiclass classification tasks. As can be seen from Tables~\ref{tab2_acc} and~\ref{tab3_acc}, our proposed DCA-CRN method outperforms the competing methods in almost all classification tasks. 

For instance, our proposed method yields the accuracy of 87.8\% and 97.5\% for eMCI vs. NC classification and AD vs. NC classification. In contrast, the best accuracies obtained by the competing methods are 86.3\% and 92.8\%, respectively. For the challenging AD vs. lMCI vs. eMCI vs. NC classification task, our proposed method achieves the overall best accuracy of 69.6\%, while the second-best overall accuracy of the four competing methods is 67.4\%. Results suggest the effectiveness of our proposed method in AD classification with rs-fMRI data.

\begin{figure}[tbh]
	\begin{center}
		\centering
		\centerline{\includegraphics[width=0.5\textwidth]{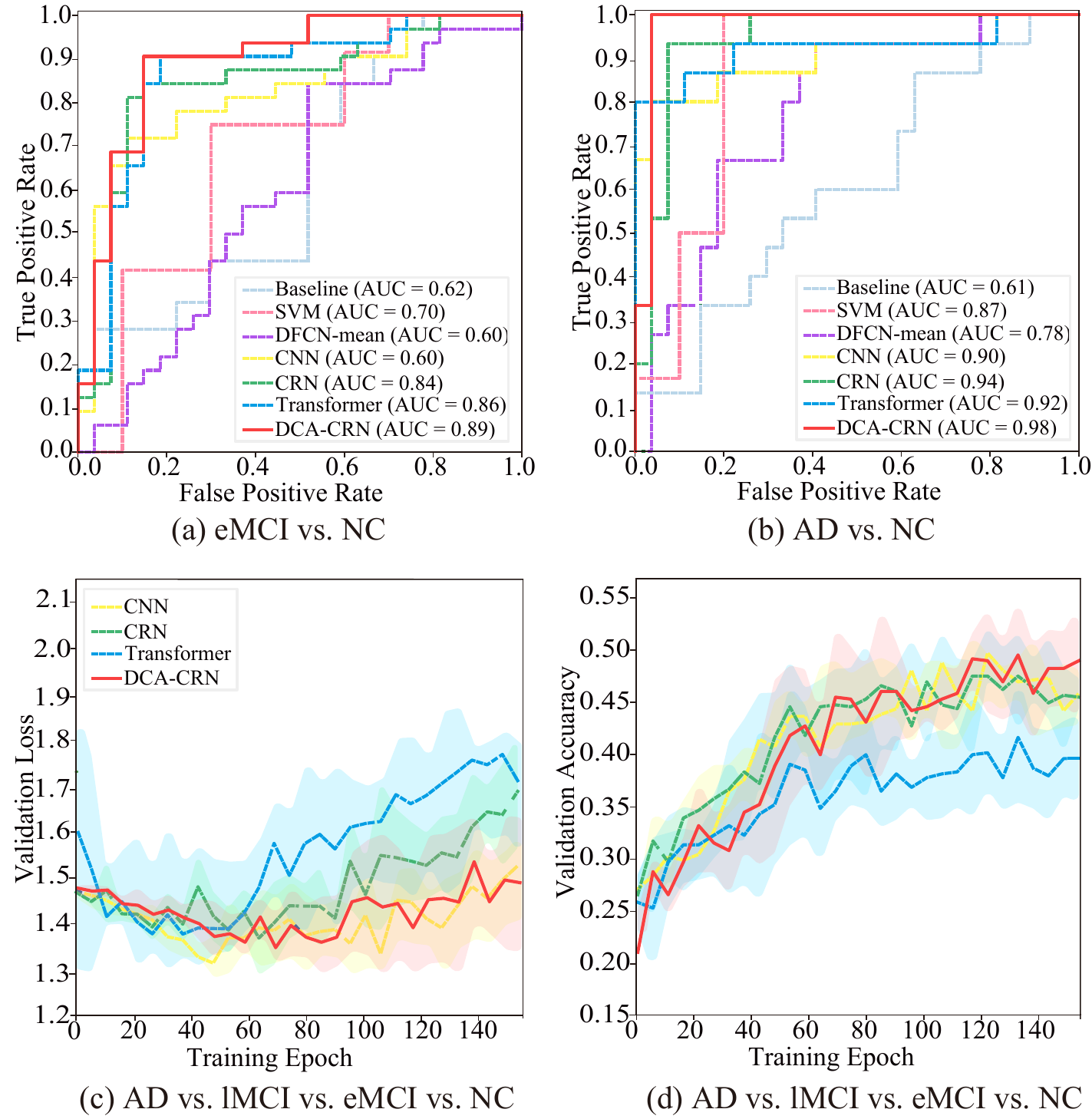}}
		\caption{(a), (b): ROC curves achieved by seven different methods in eMCI vs.NC classification (left) and AD vs. NC classification (right). (c), (d): 5-fold average validation losses and accuracies of the CNN, CRN, Transformer, and DCA-CRN methods with the first 150 epochs for AD vs. lMCI vs. eMCI vs. NC classification task. }
		\label{fig_performance}
	\end{center}
\end{figure}

In addition, From Table~\ref{tab2_acc}, \ref{tab3_acc} and Fig.~\ref{fig_performance}(a, b), we provide some interesting new findings. First, methods based on dFCNs (i.e., M2TFS, CNN, CRN, and Transformer) are generally superior to methods based on sFCN (i.e., baseline and SVM), suggesting that dynamic changes of the rs-fMRI time series can provide useful information for a better understanding of AD pathology. Second, compared with traditional SVM-based methods (i.e., baseline, SVM, and DFCN-mean), methods using the CNN or Transformer based structures can achieve better accuracy, sensitivity, and specificity performance. This shows that deep learning methods can capture the potential properties of the brain network, so they can be applied to various tasks of brain network analysis. Third, the performance of the Transformer and DCA-CRN methods is further improved, which proves the advantage of reconstructing high-order sequential features from dFCNs based on diversified covert attention patterns. Compared with the Transformer, DCA-CRN boasts better performance stability of an effective structural combination of the CNN, LSTM, and self-attention mechanism.

On the other hand, Fig.~\ref{fig_performance}(c, d) plots the validation loss and accuracy curves of validation subjects in each cross-validation fold for the AD vs. lMCI vs. eMCI vs. NC classification task. It can be seen from Fig.~\ref{fig_performance}(c, d) that the CNN, CRN, and DCA-CRN correspond to a lower and smoother decrease in validation loss compared with the Transformer. 
Additionally, as a novel deep learning method that combines the advantages of the CRN and Transformer, DCA-CRN inherits decent learning performance, which can be seen in Fig.~\ref{fig_performance}(c, d). 
This suggests that, compared with the Transformer, DCA-CRN can improve the generalization ability to unseen datasets by reconstructing high-order sequence features from dFCNs based on diversified covert attention patterns.

\subsection{Generalization contrast}
\begin{figure*}[tbh]
	\begin{center}
		\centering
		\centerline{\includegraphics[width=1.0\textwidth]{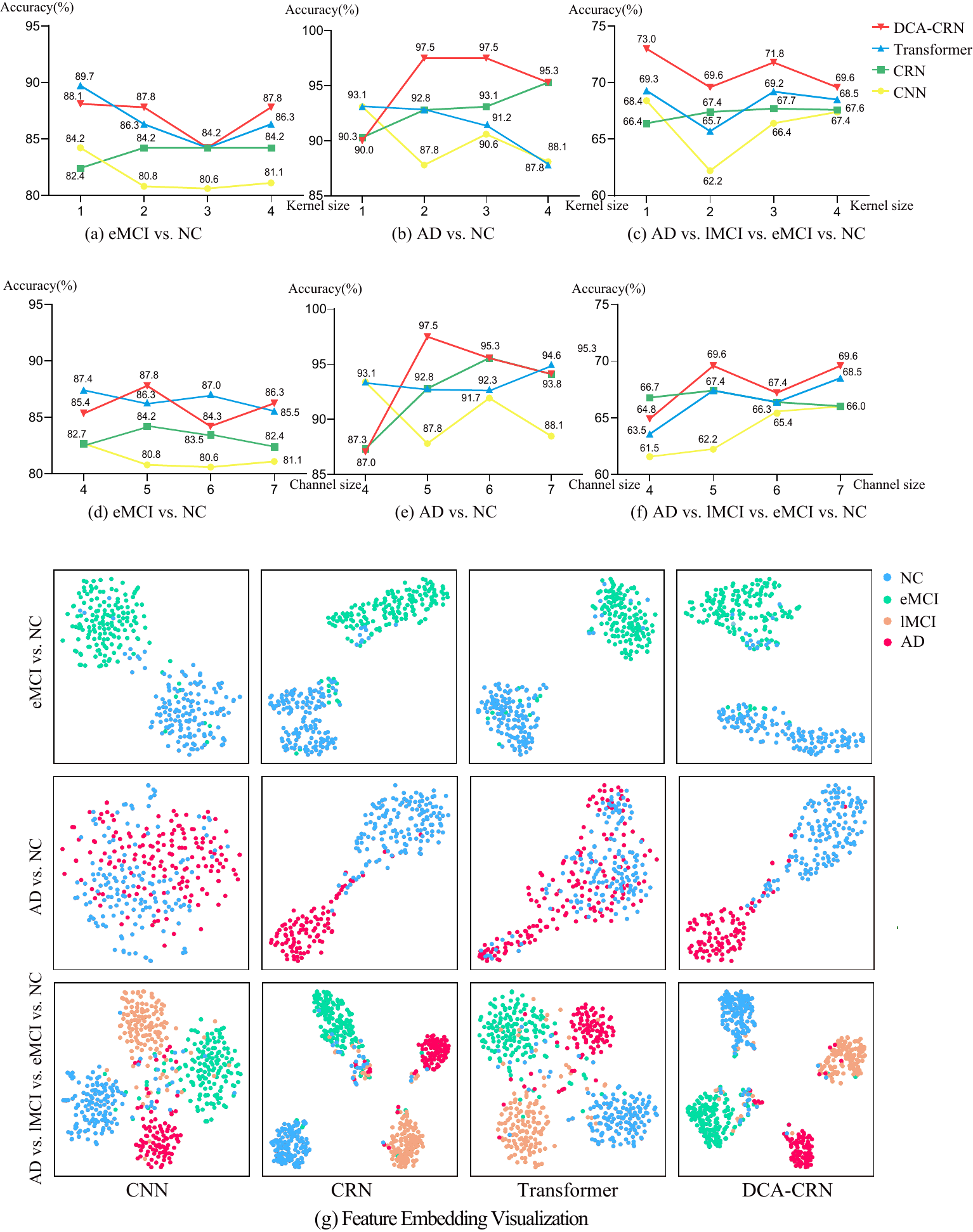}}
		\caption{(a), (b), (c): The classification accuracy of the CNN, CRN, and DCA-CRN method under different convolution kernel sizes on tasks of the eMCIvs. NC classification, the AD vs. NC classification, and the AD vs.lMCI vs. eMCI vs. NC classification. (d), (e), (f): The influence of channel sizes of our proposed DCA-CRN method in the corresponding classification tasks. (g): The t-SNE visualization for output features of the feature extraction stage for different classification tasks.}
		\label{fig_constract}
	\end{center}
\end{figure*}

To further verify the generalization ability of our DCA-CRN framework, we use four different convolution kernel and channel sizes for the layer $\mathbf{Con}1$ to check the testing accuracy of our method, as presented in Fig.~\ref{fig_constract}(a, b, c, d, e, f). First, as shown in Fig.~\ref{fig_constract}(a, b, c), our DCA-CRN method maintains leading performance under four different kernel size settings. Although the Transformer exhibits good test accuracy, its performance stability is relatively weak compared with the CRN and DCA-CRN methods. Second, as shown in Fig.~\ref{fig_constract}(d, e, f), DCA-CRN and the Transformer perform better test accuracy compared with the CNN and CRN, indicating the importance of using the self-attention mechanism to model diversified covert attention patterns for improving method performance.

We visualize output features of the feature extraction stage in the CNN, CRN, Transformer, and DCA-CRN via the t-SNE algorithm in Fig.~\ref{fig_constract}(g), where data for visualization belongs to the 5-fold cross-validation. Fig.~\ref{fig_constract}(g) shows that CNN-trained and Transformer-trained features corresponding to different classes are not well separated, especially for the AD vs. NC classification task.
We also observe that features trained with the CRN present a distribution that is locally clustered and globally separated; nevertheless, features trained with DCA-CRN are more uniformly distributed, which facilitates our method to be more sensitive to hard samples and boosts the classification performance \citep{wang2021understanding}.

\subsection{Discriminative power of learned features}
\label{sec:discussion}

\begin{figure*}[tph]
	\begin{center}
		\centering
		\includegraphics[width=1.0\textwidth]{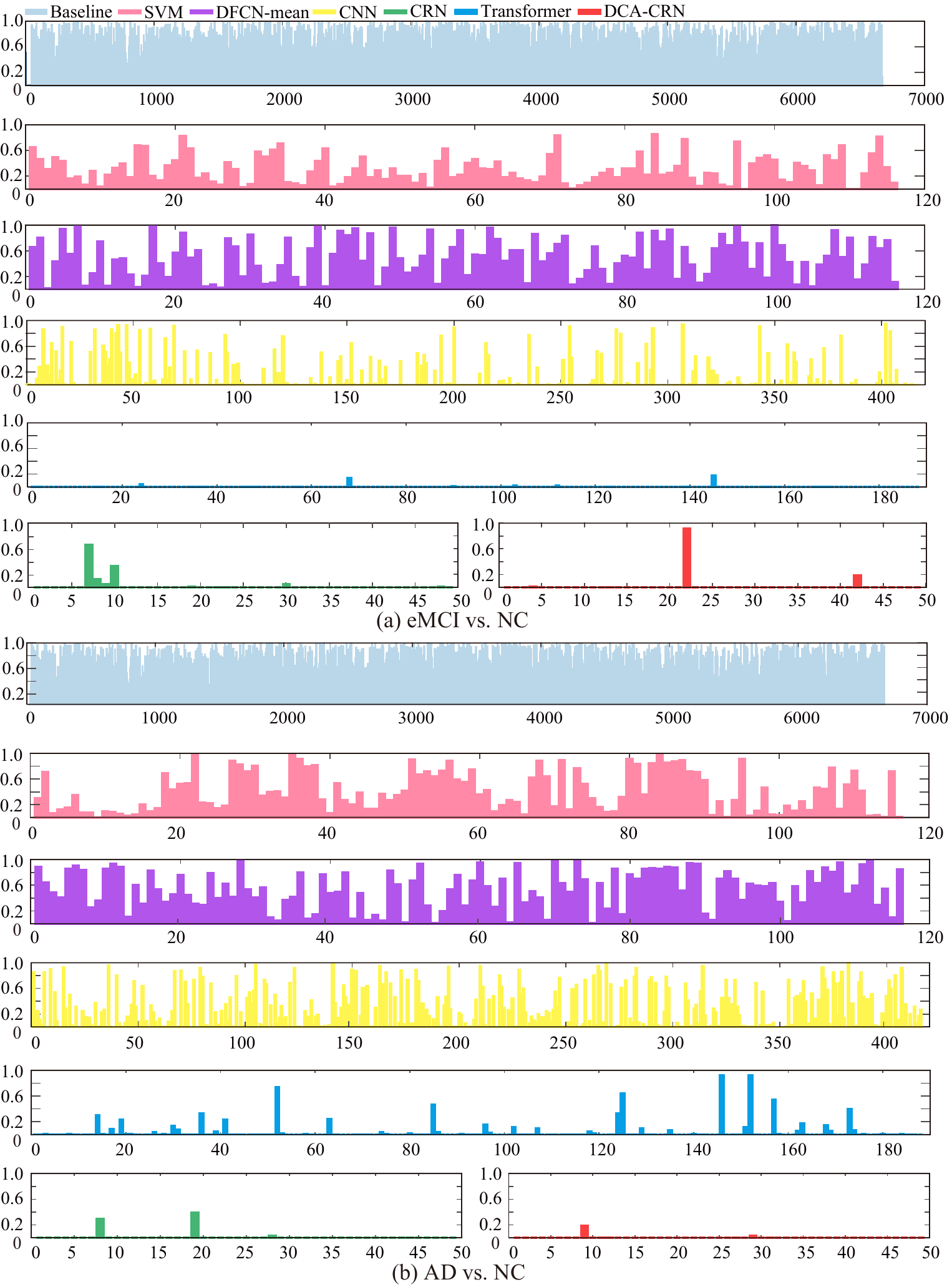}
		\caption{The learned discriminative power of features in (a): eMCI vs. NC classification and (b): AD vs. NC classification. }
		\label{fig_feature_compare}
	\end{center}
\end{figure*}

This section studies the discriminative power of features learned from our proposed DCA-CRN method. Precisely, we extract the high-order sequence features from the model, which correspond to the output of the feature extraction stage in the CNN, CRN, Transformer, and DCA-CRN methods, respectively. For traditional methods without using deep learning techniques, such as Baseline, CC, M$^{2}$TFS, we also extract the mapping sequence features from the dFCNs concerning time variation. There are a total of 6,670 connectivity strength features in the baseline method, 116 features in SVM-based methods (i.e., CC and M$^{2}$TFS), 416 features in the CNN, 48 features in the CRN, 187 features in the Transformer and 48 features in our DCA-CRN, respectively. At last, we use the standard t-test to calculate the discriminative power of all sequence features in the eMCI vs. NC and AD vs. NC group, with p-values shown in Fig.~\ref{fig_feature_compare}, respectively.

From Fig.~\ref{fig_feature_compare}, we can observe that the p-values of the features learned by the CRN and DCA-CRN methods are mainly close to 0 (i.e., very sparsity) compared with other methods. It implies that both methods can distinguish eMCI and AD from NCs. In addition, the p-values of our proposed DCA-CRN method are more sparse than those of the CRN method, indicating that the features learned by the DCA-CRN method are more discriminative than those of the CRN method.

\section{Discussion}
\subsection{3D visualization of discriminative FCNs and medical analysis}
\begin{figure*}[tph]
	\begin{center}
		\centering
		\includegraphics[width=1.0\textwidth]{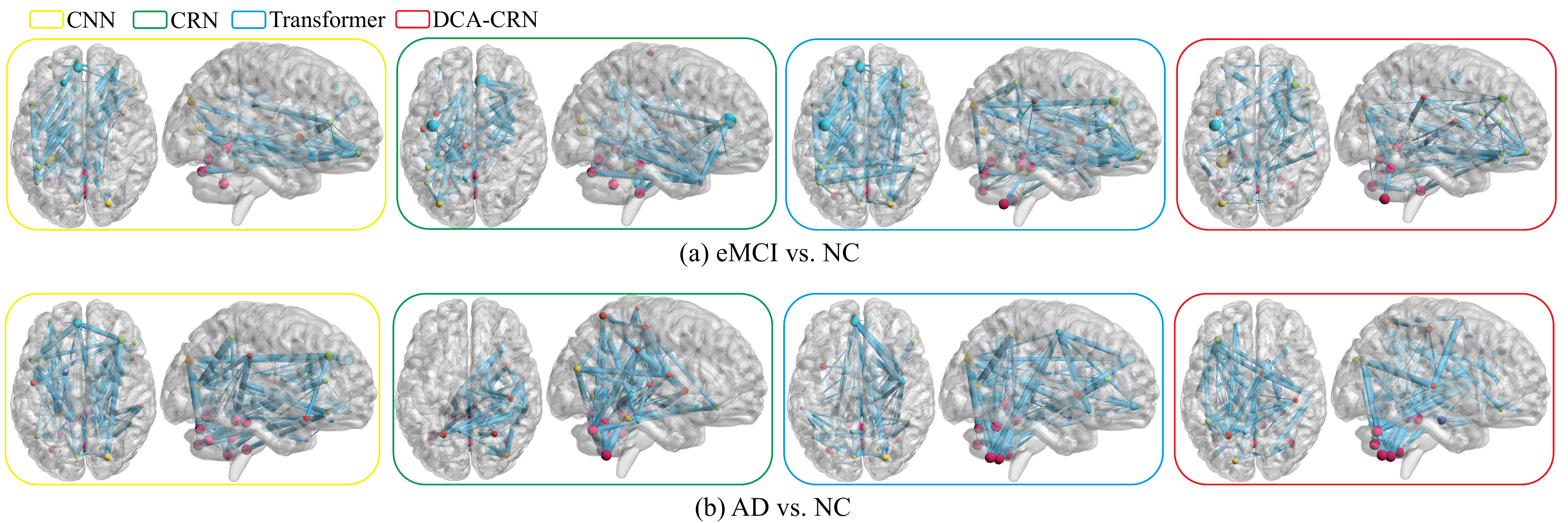}
		\caption{3D visualization of the top important brain regions and corresponding discriminative functional brain connectivity networks for (a) eMCI vs. NC and (b) AD vs. NC classification.}
		\label{fig_connectivity_compare}
	\end{center}
\end{figure*}

AD is a complex brain disease that involves complex dynamic changes of FCNs at different stages. Nevertheless, there are vast differences in the discriminative FCNs proposed by existing deep learning approaches due to the use of different databases or different brain region templates, etc. Therefore, we use our proposed method to explore discriminative FCNs and corresponding brain regions compared with the CNN, CRN, and Transformer methods under identical conditions such as dataset partitioning.
First, we extract high-order sequence features after the first convolution layer in the CNN, CRN, Transformer, and DCA-CRN and test them in a separate t-test ($\mathbf{p}<0.05$) significance groups by channel, to explore top important brain regions which contributed to the specific classification tasks, respectively. Then, we select the top important brain regions and test the connectivity among them in a separate t-test ($\mathbf{p}<0.05$) significance groups by channel to explore the discriminative FCNs for the specific classification tasks, respectively.
Fig.~\ref{fig_connectivity_compare} depict the top important brain regions and corresponding discriminative FCNs for (a) eMCI vs. NC group and (b) AD vs. NC group, separately. 

By comparing DCA-CRN with other methods, we obtain some meaningful findings on using deep learning methods for exploring important brain regions associated with  AD, as shown in Fig.~\ref{fig_connectivity_compare}. 
For example, cerebellar regions also appear in the important brain regions and corresponding discriminative FCNs. Furthermore, it is worth emphasizing that the connectivity between cerebellar and cerebral regions is significantly enhanced in AD vs. NC, compared with eMCI vs. NC.

For eMCI vs. NC classification, the most brain regions and corresponding discriminative FCNs our proposed DCA-CRN method explored can be found in the CNN, CRN, and Transformer methods. These brain regions have been shown to have significant associations with AD, including not only cerebral regions such as the right middle frontal gyrus (MFG.R) \citep{chen2022spatially}, the left middle orbitofrontal cortex (ORBmid.L) \citep{kumfor2013orbitofrontal}, the left triangular inferior frontal gyrus (IFGtriang.L) \citep{penniello1995pet}, the left rectus gyrus (REC.L) \citep{ballmaier2004anterior}, the left calcarine cortex (CAL.L) \citep{wright2007functional}, the right supramarginal gyrus (SMG.R) \citep{mcdonough2020risk}, but also cerebellar regions such as the lobule IV, V of vermis (Vermis45) \citep{schmahmann2016cerebellum}, the lobule VI of vermis (Vermis6) \citep{gellersen2017cerebellar,schmahmann2016cerebellum}. Encouragingly, the role of the cerebellum in cognitive function has also been extensively studied from anatomical, clinical, and functional perspectives over the last few decades, and new evidence suggests that the cerebellum also has a significant contribution to make in cognitive function in AD \citep{bernard2022don,jacobs2018cerebellum,gellersen2017cerebellar}. Compared with the CNN and CRN methods, the Transformer and our proposed method find new cerebellar regions and the corresponding discriminative brain function connectivity subnetworks associated with the eMCI stage. Sure of these brain regions have also been shown to have high underlying associations with AD in recent studies, including the left crus I of cerebellar hemisphere (CRBLCrus1.L) \citep{toniolo2018patterns}, the left crus II of cerebellar hemisphere (CRBLCrus2.L) \citep{chen2022cerebellar}.

For AD vs. NC classification, discriminative FCNs explored by our proposed method retain similar to that explored by the CNN, CRN, and Transformer methods. Important brain regions which have been proved to be correlated with AD include IFGtriang.L \citep{penniello1995pet}, CRBLCrus1.L \citep{toniolo2018patterns}, Vermis6 \citep{gellersen2017cerebellar,schmahmann2016cerebellum}. By comparing changes in discriminative FCNs from the eMCI vs. NC classification to the AD vs. NC classification, we find that posterior cerebellar regions, such as the right crus II of cerebellar hemisphere (CRBLCrus2.R) and the right lobule VIIB of cerebellar hemisphere (CRBL7b.R), contribute more to AD than eMCI. It has been studied that the anterior cerebellum contributes more to mild cognitive impairment (MCI), and the posterior cerebellum contributes more to the mild and moderate stages of AD \citep{bruchhage2020machine}. Thus, our finding keeps consistent with the study by \citep{bruchhage2020machine} and demonstrates the cerebellum's potential role in cognition for AD. 

\begin{figure*}[tph]
	\begin{center}
		\centering
		\includegraphics[width=1.0\textwidth]{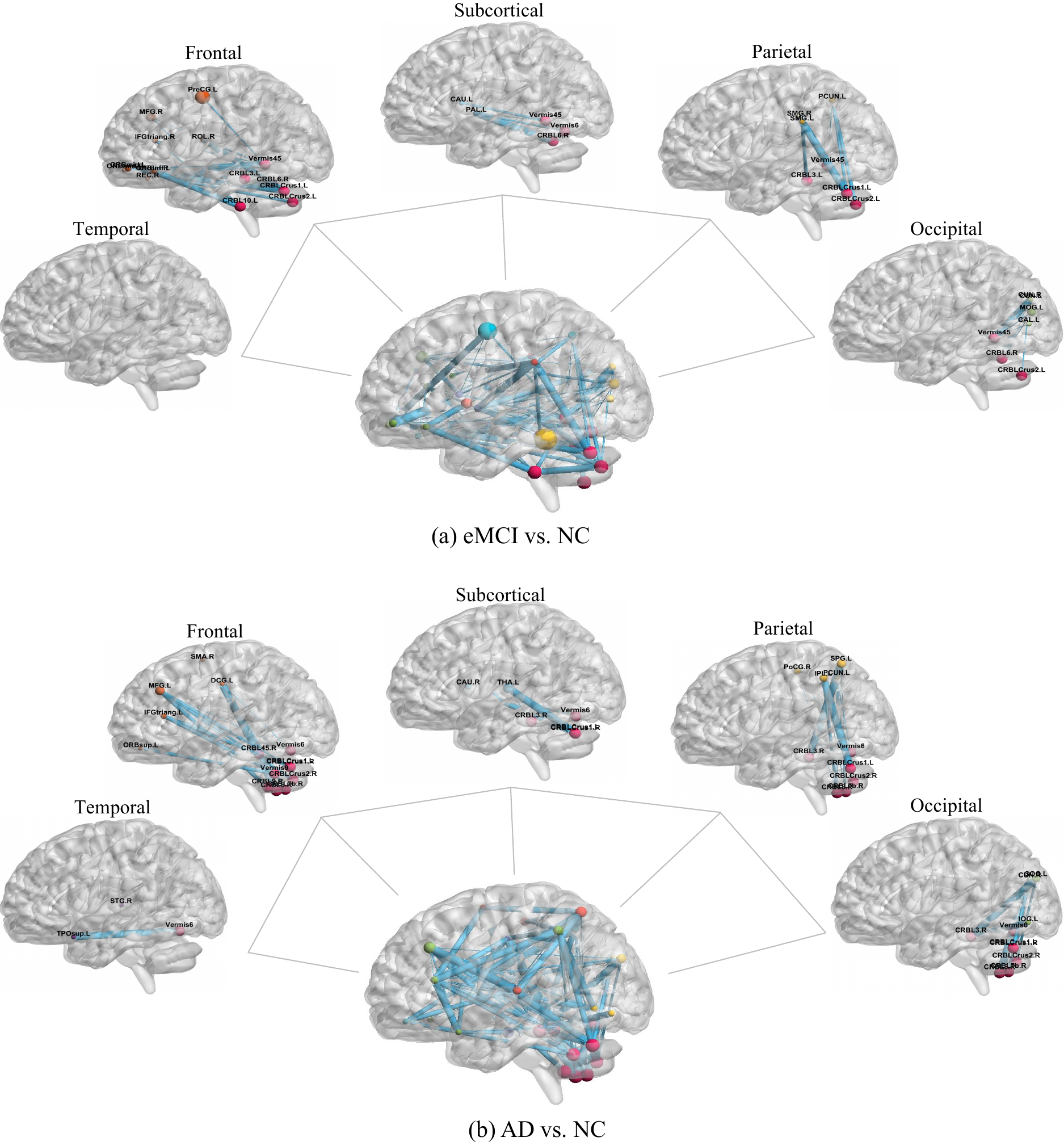}
		\caption{3D visualization of discriminative functional connectivity subnetworks among different cerebral regions and cerebellar regions corresponding to eMCI vs. NC and AD vs. NC, respectively.}
		\label{connectivity_sa_crn}
	\end{center}
\end{figure*}

\begin{figure*}[tph]
	\begin{center}
		\centering
		\includegraphics[width=1.0\textwidth]{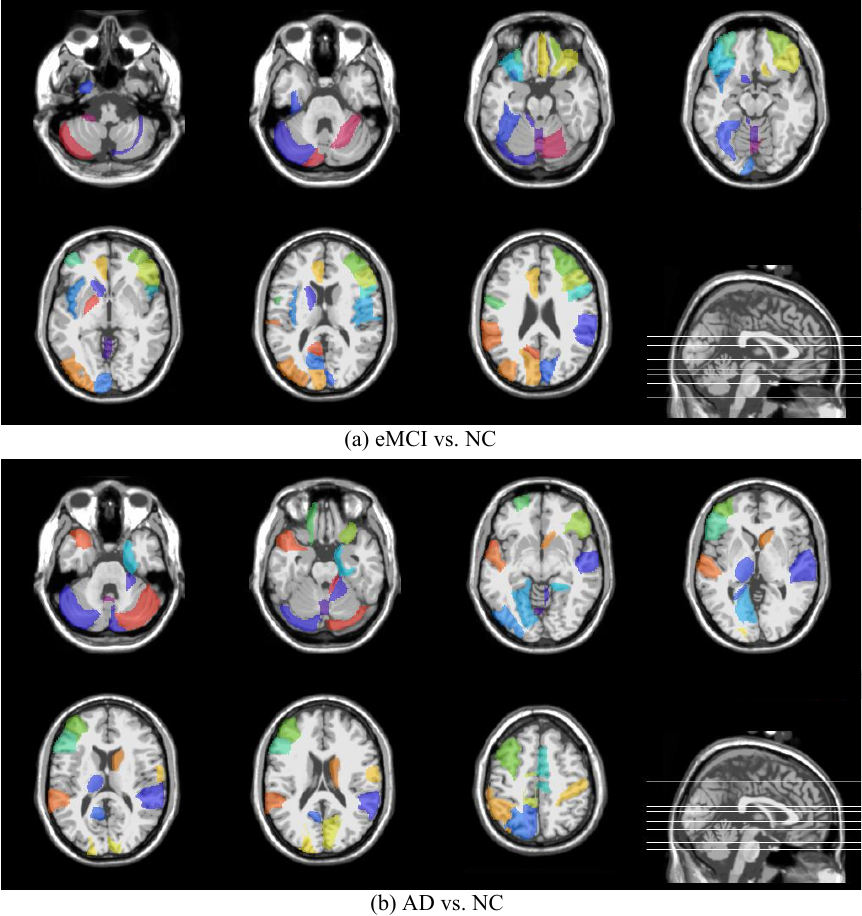}
		\caption{Discriminative brain regions identified by the proposed method in (a) eMCI vs. NC classification and (b) AD vs. NC classification.}
		\label{fig_ROI_SA_CRN}
	\end{center}
\end{figure*}

Furthermore, we visualize and analyze discriminative functional connectivity subnetworks among the cerebellar and cerebral regions corresponding to eMCI vs. NC and AD vs. NC using our proposed method in Fig.~\ref{connectivity_sa_crn}, respectively. First, in comparison to the eMCI stage, the functional connectivity between the temporal and occipital lobe regions and the cerebellar region exhibited further enhancement between AD and NCs, as evidenced by brain regions such as the right cuneus (CUN.R), the left superior temporal pole (TPOsup.L), and the right superior temporal gyrus (STG.R), suggesting a further weakening AD patients' abilities in spatial perception, emotional understanding and control, and social cognition \citep{love2016cerebrovascular,braak1989alzheimer}. Second, in contrast to NCs, the functional connectivity between subcortical regions and cerebellar regions is found to be discriminative for eMCI and AD patients, as exemplified by regions such as the right caudate (CAU.R), the left pallidum (PAL.L), and the right thalamus (THA.R). That indicates that eMCI and AD patients experience varying degrees of difficulty in attention regulation, motor execution control, and habit formation, among other functions. \citep{landin2017disease,kavcic2003attentional, mcduff1985subcortical}. As a supplement, Fig.~\ref{fig_ROI_SA_CRN} shows the sagittal plane visualization of the important brain region we have selected.

\subsection{Inspecting the $\mathbf{DCA}$ layer}
\begin{figure*}[tph]
	\begin{center}
		\centering
		\includegraphics[width=1.0\textwidth]{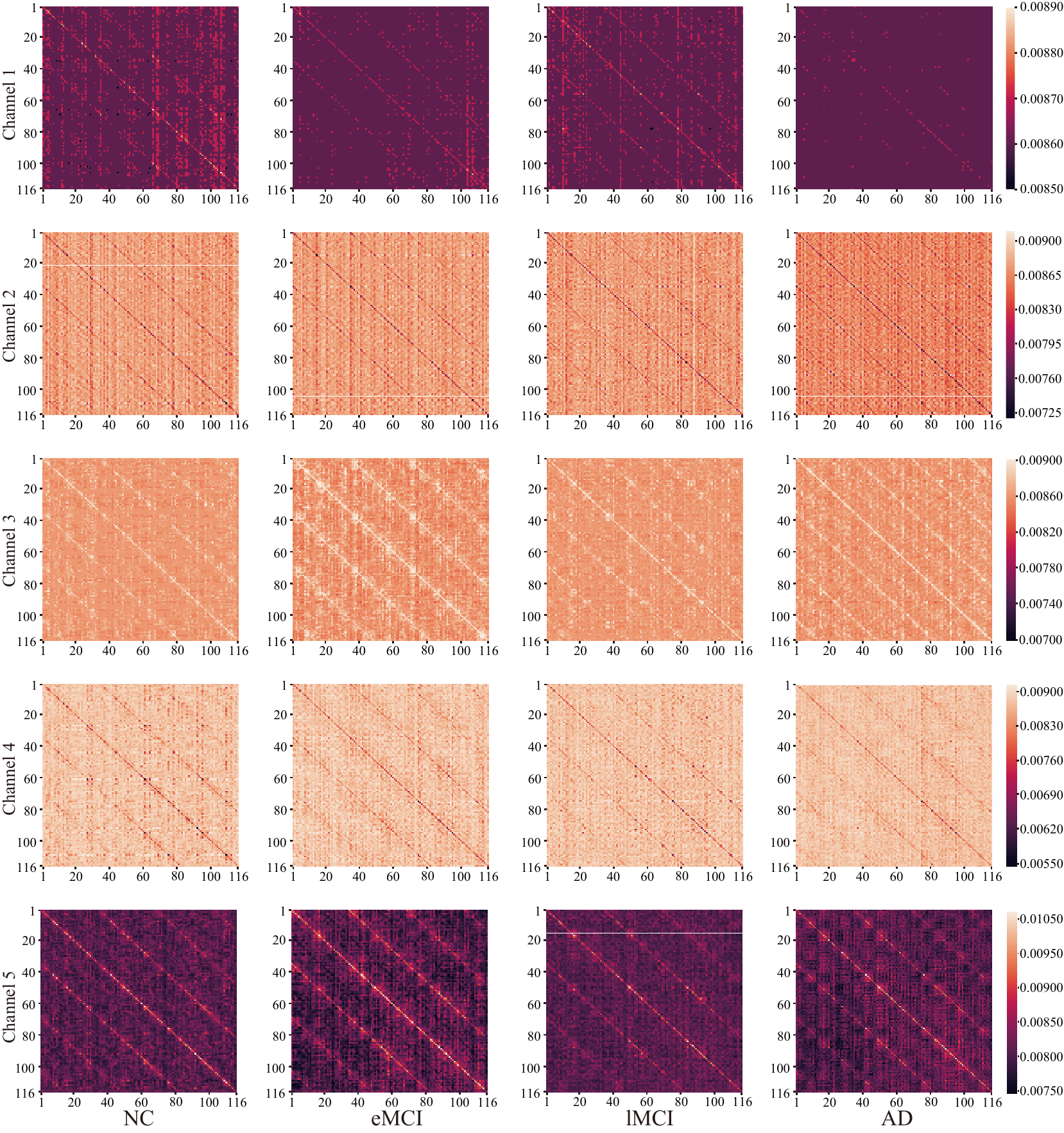}
		\caption{Heatmap visualization of attention score vectors for subjects from different AD stages. The $\mathbf{DCA}$ layer has 5 channels, and every row of each attention score vector indicates correlations from all brain regions to the specified brain region.}
		\label{fig_score}
	\end{center}
\end{figure*}

To explore internal details inside the $\mathbf{DCA}$ layer, we first select rs-fMRI data of subjects with the highest ranked model prediction accuracy under four categories as inputs for our DCA-CRN model. Next, we extract attention score vectors $\mathbf{P}$ from the $\mathbf{DCA}$ layer and perform heat maps, where $\mathbf{P}_{\mathbf{ij}}$ indicates dynamic correlation scores between the spatio-temporal features of brain regions $\mathbf{i}$ and $\mathbf{j}$, as shown in Fig.~\ref{fig_score}.

By observing Fig.~\ref{fig_score}, we can obtain the following findings: First, attention score vectors of AD versus eMCI and lMCI yield inconsistent patterns. In channel 1, eMCI, lMCI, and AD are more sparsely highlighted in the heatmap compared with the NC level. That implies that the trained model focuses on different brain regions for diseased individuals than for normal ones when making predictions. Moreover, the correlation intensity between cerebellum and cerebrum is higher for NCs to MCIs and minimal for AD, where MCI refers to eMCI and lMCI \citep{shireby2022dna}. For other channels, the differences across subjects of diverse types are not evident. In contrast, we infer that the model learns more about common covert attention patterns of correlations in these channels. Also, attention score vectors after Softmax show little value difference between each brain region, most concentrated at the average level (0.0086). One possible explanation is that the model's inputs range from 0 to 1 during the forward propagation of the network, and feature values are also small under the control of BN, LN, and RELU activation functions; thus, even a tiny variability may have a qualitative impact on the output. Similarly, within several heatmaps, localized clusterings of highlighted areas appear because adjacent brain regions may function in closer proximity and influence each other. 

Overall, the DCA mechanism improves model prediction performance by variably learning higher-order feature correlations among different brain regions for diversified types of subjects, as well as mining more complex underlying higher-order and high-level brain functional connectivity information.

\subsection{Additional validation on the ADHD classification}
\label{sec:adhd}

ADHD is a common neurodevelopmental disorder characterized by inattention, hyperactivity, and impulsivity. Clinical diagnosis of ADHD still faces some difficulties and challenges, such as diverse symptom presentations, the lack of specific biomarkers, and age dependency. Deep learning-based brain disease classification methods have also conducted experiments and analyses on this disease \citep{zhang2022diffusion,liu2022enhanced}. 

Thus, we collect rs-fMRI data for $216$ subjects, including $98$ NCs and $31$ ADHDs from the ADHD-200 dataset, as shown in Table~\ref{tab1_subjects}. The ADHD-200 dataset from the NYU website was processed using the Athena pipeline~\footnote{https://www.nitrc.org/plugins/mwiki/index.php/neurobureau:Athena}. The initial step involved removing the first four EPI volumes to balance the signal. That is followed by slice timing, head motion correction, and co-registration of the echo-planar average image to the MNI template space with a resolution of 4×4×4. Then, the average time series for specific brain regions were obtained from the gray matter time series using the AAL-116 atlas and used as the input. 

\begin{table}[!tbp]	
	\scriptsize
	\renewcommand{\arraystretch}{1.2}
	\caption{Performance of seven methods in  ADHD vs. NC classification task. ACC=Accuracy.} 
	\begin{tabular*}{0.48\textwidth}{@{\extracolsep{\fill}}l ccc}
		\toprule
		\multirow{2}{*}{~Method}&\multicolumn{3}{c}{ADHD vs. NC (\%)} \\
		\cline{2-4}
		& ACC & SPE  & SEN \\
		\hline
		~Baseline & $63.0$ & $68.6$ & $56.1$ \\
		
		~SVM & $75.6$ & $74.9$ & $77.2$ \\
		
		~DFCN-mean & $67.7$ & $71.9$ & $61.7$ \\
		
		~CNN  & $79.7$ & $75.2$ & $85.2$ \\
		
		~CRN  & $85.0$ & $78.8$ & $90.0$\\
		
		~Transformer & $83.0$ & $78.3$ & $87.0$ \\
		
		~DCA-CRN(OURS)   & \textbf{$87.0$} & $86.9$ & \textbf{$87.0$} \\	
		\bottomrule
	\end{tabular*}
	\label{tab4_acc}
\end{table} 

As a result, Table \ref{tab4_acc} shows the performance results of different methods on the ADHD vs. NC classification task, where DCA-CRN achieved leading scores in ACC, SPE, and SEN metrics. As shown in Table \ref{tab4_acc}, deep learning-based methods have clear performance advantages over traditional group testing and machine learning methods, with better ADHD diagnostic capabilities. Notably, DCA-CRN exhibits more balanced stability in SPE and SEN, indicating that DCA-CRN has high prediction accuracy and generalization performance.

\subsection{Limitations and future Work}
Although DCA-CRN has shown good learning and generalization performance in different brain disease classification tasks, some shortcomings in this study can be further improved and perfected: First, collect data from multiple AD databases to increase the amount of experimental data. Second, consider multimodal information and extend the idea of diversified covert attention patterns in DCA-CRN to the latest multimodal deep learning models applied to brain disease classification. Third, improve the interpretability of the covert attention patterns, not just the interpretability of framework design. 

In the future work, we would further explore the potential effect of diversified covert attention patterns on rs-fMRI data acquisition for subjects with different brain diseases, and investigate the interpretability of differences in covert attention patterns. Additionally, we would attempt to apply the insight of covert attention patterns to other medical imaging data beyond rs-fMRI.

\section{Conclusion}
In this study, we propose a novel convolutional recurrent neural network model, DCA-CRN, which reconstructs high-order sequence features of dFCNs based on diversified covert attention patterns and combines the advantages of the CRN and Transformer. First, we conduct a series of comparative experiments on the ADNI and ADHD-200 datasets to verify the superior prediction and generalization performance of DCA-CRN compared with the CNN, CRN, and Transformer. Second, by visualizing the feature distributions of different types of AD subject groups corresponding to the CNN, CRN, Transformer, and DCA-CRN methods, we find that the feature distribution trained by DCA-CRN is more uniform, making our method more sensitive to hard samples and improving classification performance. Third, most of the important brain regions found by DCA-CRN in the NC vs. eMCI and NC vs. AD classification tasks are consistent with existing medical research. In addition, DCA-CRN and the Transformer find more cerebellar regions related to eMCI, AD, and corresponding discriminant brain functional connectivity subnetworks. We argue that this is because both utilize the self-attention mechanism to model covert attention patterns of dFCNs within brain information flow. These results further highlight the importance of cerebellar research for early prediction and diagnosis of AD. Lastly, we hope that this study will be helpful to the research of AD and that the diagnosis challenge of AD can be solved in the not-too-distant future.  

\section{Acknowledgement}
Z. Zhang, B. Jie, Z. Wang, J. Zhou and Y. Yang are supported in part by NSFC (Nos.61976006, 61573023, 61902003), Anhui-NSFC (Nos.1708085MF145, 1808085MF171) and AHNU-FOYHE (No.gxyqZD2017010).

%%Harvard
\bibliographystyle{model2-names.bst}\biboptions{authoryear}
\bibliography{refs}

\begin{thebibliography}{76}
\expandafter\ifx\csname natexlab\endcsname\relax\def\natexlab#1{#1}\fi
\providecommand{\url}[1]{\texttt{#1}}
\providecommand{\href}[2]{#2}
\providecommand{\path}[1]{#1}
\providecommand{\DOIprefix}{doi:}
\providecommand{\ArXivprefix}{arXiv:}
\providecommand{\URLprefix}{URL: }
\providecommand{\Pubmedprefix}{pmid:}
\providecommand{\doi}[1]{\href{http://dx.doi.org/#1}{\path{#1}}}
\providecommand{\Pubmed}[1]{\href{pmid:#1}{\path{#1}}}
\providecommand{\bibinfo}[2]{#2}
\ifx\xfnm\relax \def\xfnm[#1]{\unskip,\space#1}\fi
%Type = Article
\bibitem[{Aisen et~al.(2022)Aisen, Jimenez-Maggiora, Rafii, Walter and
  Raman}]{aisen2022early}
\bibinfo{author}{Aisen, P.S.}, \bibinfo{author}{Jimenez-Maggiora, G.A.},
  \bibinfo{author}{Rafii, M.S.}, \bibinfo{author}{Walter, S.},
  \bibinfo{author}{Raman, R.}, \bibinfo{year}{2022}.
\newblock \bibinfo{title}{Early-stage alzheimer disease: getting trial-ready}.
\newblock \bibinfo{journal}{Nature Reviews Neurology} \bibinfo{volume}{18},
  \bibinfo{pages}{389--399}.
%Type = Article
\bibitem[{Ballmaier et~al.(2004)Ballmaier, Toga, Blanton, Sowell, Lavretsky,
  Peterson, Pham and Kumar}]{ballmaier2004anterior}
\bibinfo{author}{Ballmaier, M.}, \bibinfo{author}{Toga, A.W.},
  \bibinfo{author}{Blanton, R.E.}, \bibinfo{author}{Sowell, E.R.},
  \bibinfo{author}{Lavretsky, H.}, \bibinfo{author}{Peterson, J.},
  \bibinfo{author}{Pham, D.}, \bibinfo{author}{Kumar, A.},
  \bibinfo{year}{2004}.
\newblock \bibinfo{title}{Anterior cingulate, gyrus rectus, and orbitofrontal
  abnormalities in elderly depressed patients: an mri-based parcellation of the
  prefrontal cortex}.
\newblock \bibinfo{journal}{American Journal of Psychiatry}
  \bibinfo{volume}{161}, \bibinfo{pages}{99--108}.
%Type = Article
\bibitem[{Bernard(2022)}]{bernard2022don}
\bibinfo{author}{Bernard, J.A.}, \bibinfo{year}{2022}.
\newblock \bibinfo{title}{Don’t forget the little brain: a framework for
  incorporating the cerebellum into the understanding of cognitive aging}.
\newblock \bibinfo{journal}{Neuroscience \& Biobehavioral Reviews}
  \bibinfo{volume}{137}, \bibinfo{pages}{104639}.
%Type = Article
\bibitem[{Bessadok et~al.(2022)Bessadok, Mahjoub and Rekik}]{bessadok2022graph}
\bibinfo{author}{Bessadok, A.}, \bibinfo{author}{Mahjoub, M.A.},
  \bibinfo{author}{Rekik, I.}, \bibinfo{year}{2022}.
\newblock \bibinfo{title}{Graph neural networks in network neuroscience}.
\newblock \bibinfo{journal}{IEEE Transactions on Pattern Analysis and Machine
  Intelligence} \bibinfo{volume}{45}, \bibinfo{pages}{5833--5848}.
%Type = Article
\bibitem[{Bhatti et~al.(2023)Bhatti, Tang, Wu, Marjan and
  Hussain}]{bhatti2023deep}
\bibinfo{author}{Bhatti, U.A.}, \bibinfo{author}{Tang, H.},
  \bibinfo{author}{Wu, G.}, \bibinfo{author}{Marjan, S.},
  \bibinfo{author}{Hussain, A.}, \bibinfo{year}{2023}.
\newblock \bibinfo{title}{Deep learning with graph convolutional networks: An
  overview and latest applications in computational intelligence}.
\newblock \bibinfo{journal}{International Journal of Intelligent Systems}
  \bibinfo{volume}{2023}, \bibinfo{pages}{1--28}.
%Type = Article
\bibitem[{Bi et~al.(2023)Bi, Chen, Jiang, Luo, Zhou, Xing, Xu, Liu and
  Liu}]{bi2023community}
\bibinfo{author}{Bi, X.A.}, \bibinfo{author}{Chen, K.}, \bibinfo{author}{Jiang,
  S.}, \bibinfo{author}{Luo, S.}, \bibinfo{author}{Zhou, W.},
  \bibinfo{author}{Xing, Z.}, \bibinfo{author}{Xu, L.}, \bibinfo{author}{Liu,
  Z.}, \bibinfo{author}{Liu, T.}, \bibinfo{year}{2023}.
\newblock \bibinfo{title}{Community graph convolution neural network for
  alzheimer’s disease classification and pathogenetic factors
  identification}.
\newblock \bibinfo{journal}{IEEE Transactions on Neural Networks and Learning
  Systems} .
%Type = Article
\bibitem[{Braak et~al.(1989)Braak, Braak and Kalus}]{braak1989alzheimer}
\bibinfo{author}{Braak, H.}, \bibinfo{author}{Braak, E.},
  \bibinfo{author}{Kalus, P.}, \bibinfo{year}{1989}.
\newblock \bibinfo{title}{Alzheimer's disease: areal and laminar pathology in
  the occipital isocortex}.
\newblock \bibinfo{journal}{Acta neuropathologica} \bibinfo{volume}{77},
  \bibinfo{pages}{494--506}.
%Type = Article
\bibitem[{Bruchhage et~al.(2020)Bruchhage, Correia, Malloy, Salloway and
  Deoni}]{bruchhage2020machine}
\bibinfo{author}{Bruchhage, M.M.}, \bibinfo{author}{Correia, S.},
  \bibinfo{author}{Malloy, P.}, \bibinfo{author}{Salloway, S.},
  \bibinfo{author}{Deoni, S.}, \bibinfo{year}{2020}.
\newblock \bibinfo{title}{Machine learning classification identifies cerebellar
  contributions to early and moderate cognitive decline in alzheimer?s
  disease}.
\newblock \bibinfo{journal}{Frontiers in Aging Neuroscience}
  \bibinfo{volume}{12}, \bibinfo{pages}{524024}.
%Type = Article
\bibitem[{Chen et~al.(2011)Chen, Ward, Xie, Li, Wu, Jones, Franczak, Antuono
  and Li}]{chen2011classification}
\bibinfo{author}{Chen, G.}, \bibinfo{author}{Ward, B.D.}, \bibinfo{author}{Xie,
  C.}, \bibinfo{author}{Li, W.}, \bibinfo{author}{Wu, Z.},
  \bibinfo{author}{Jones, J.L.}, \bibinfo{author}{Franczak, M.},
  \bibinfo{author}{Antuono, P.}, \bibinfo{author}{Li, S.J.},
  \bibinfo{year}{2011}.
\newblock \bibinfo{title}{Classification of alzheimer disease, mild cognitive
  impairment, and normal cognitive status with large-scale network analysis
  based on resting-state functional mr imaging}.
\newblock \bibinfo{journal}{Radiology} \bibinfo{volume}{259},
  \bibinfo{pages}{213--221}.
%Type = Article
\bibitem[{Chen et~al.(2022a)Chen, Chang, Li, Acosta, Li, Guo, Wang, Turkes,
  Morrison, Julian et~al.}]{chen2022spatially}
\bibinfo{author}{Chen, S.}, \bibinfo{author}{Chang, Y.}, \bibinfo{author}{Li,
  L.}, \bibinfo{author}{Acosta, D.}, \bibinfo{author}{Li, Y.},
  \bibinfo{author}{Guo, Q.}, \bibinfo{author}{Wang, C.},
  \bibinfo{author}{Turkes, E.}, \bibinfo{author}{Morrison, C.},
  \bibinfo{author}{Julian, D.}, et~al., \bibinfo{year}{2022}a.
\newblock \bibinfo{title}{Spatially resolved transcriptomics reveals genes
  associated with the vulnerability of middle temporal gyrus in alzheimer’s
  disease}.
\newblock \bibinfo{journal}{Acta Neuropathologica Communications}
  \bibinfo{volume}{10}, \bibinfo{pages}{1--24}.
%Type = Article
\bibitem[{Chen et~al.(2017a)Chen, Zhang, Lee, Shen and
  Initiative}]{chen2017hierarchical}
\bibinfo{author}{Chen, X.}, \bibinfo{author}{Zhang, H.}, \bibinfo{author}{Lee,
  S.W.}, \bibinfo{author}{Shen, D.}, \bibinfo{author}{Initiative, A.D.N.},
  \bibinfo{year}{2017}a.
\newblock \bibinfo{title}{Hierarchical high-order functional connectivity
  networks and selective feature fusion for mci classification}.
\newblock \bibinfo{journal}{Neuroinformatics} \bibinfo{volume}{15},
  \bibinfo{pages}{271--284}.
%Type = Article
\bibitem[{Chen et~al.(2017b)Chen, Zhang, Zhang, Shen, Lee and
  Shen}]{chen2017extraction}
\bibinfo{author}{Chen, X.}, \bibinfo{author}{Zhang, H.},
  \bibinfo{author}{Zhang, L.}, \bibinfo{author}{Shen, C.},
  \bibinfo{author}{Lee, S.W.}, \bibinfo{author}{Shen, D.},
  \bibinfo{year}{2017}b.
\newblock \bibinfo{title}{Extraction of dynamic functional connectivity from
  brain grey matter and white matter for mci classification}.
\newblock \bibinfo{journal}{Human brain mapping} \bibinfo{volume}{38},
  \bibinfo{pages}{5019--5034}.
%Type = Article
\bibitem[{Chen et~al.(2022b)Chen, Landin-Romero, Kumfor, Irish, Dobson-Stone,
  Kwok, Halliday, Hodges and Piguet}]{chen2022cerebellar}
\bibinfo{author}{Chen, Y.}, \bibinfo{author}{Landin-Romero, R.},
  \bibinfo{author}{Kumfor, F.}, \bibinfo{author}{Irish, M.},
  \bibinfo{author}{Dobson-Stone, C.}, \bibinfo{author}{Kwok, J.B.},
  \bibinfo{author}{Halliday, G.M.}, \bibinfo{author}{Hodges, J.R.},
  \bibinfo{author}{Piguet, O.}, \bibinfo{year}{2022}b.
\newblock \bibinfo{title}{Cerebellar integrity and contributions to cognition
  in c9orf72-mediated frontotemporal dementia}.
\newblock \bibinfo{journal}{Cortex} \bibinfo{volume}{149},
  \bibinfo{pages}{73--84}.
%Type = Article
\bibitem[{Culham et~al.(2001)Culham, Cavanagh and
  Kanwisher}]{culham2001attention}
\bibinfo{author}{Culham, J.C.}, \bibinfo{author}{Cavanagh, P.},
  \bibinfo{author}{Kanwisher, N.G.}, \bibinfo{year}{2001}.
\newblock \bibinfo{title}{Attention response functions: characterizing brain
  areas using fmri activation during parametric variations of attentional
  load}.
\newblock \bibinfo{journal}{Neuron} \bibinfo{volume}{32},
  \bibinfo{pages}{737--745}.
%Type = Article
\bibitem[{David et~al.(2022)David, Del~Giovane, Liu, Gostick, Rowe, Oboh,
  Howard and Malhotra}]{david2022cognitive}
\bibinfo{author}{David, M.C.}, \bibinfo{author}{Del~Giovane, M.},
  \bibinfo{author}{Liu, K.Y.}, \bibinfo{author}{Gostick, B.},
  \bibinfo{author}{Rowe, J.B.}, \bibinfo{author}{Oboh, I.},
  \bibinfo{author}{Howard, R.}, \bibinfo{author}{Malhotra, P.A.},
  \bibinfo{year}{2022}.
\newblock \bibinfo{title}{Cognitive and neuropsychiatric effects of
  noradrenergic treatment in alzheimer’s disease: Systematic review and
  meta-analysis}.
\newblock \bibinfo{journal}{Journal of Neurology, Neurosurgery \& Psychiatry}
  \bibinfo{volume}{93}, \bibinfo{pages}{1080--1090}.
%Type = Article
\bibitem[{Faraone et~al.(2003)Faraone, Sergeant, Gillberg and
  Biederman}]{faraone2003worldwide}
\bibinfo{author}{Faraone, S.V.}, \bibinfo{author}{Sergeant, J.},
  \bibinfo{author}{Gillberg, C.}, \bibinfo{author}{Biederman, J.},
  \bibinfo{year}{2003}.
\newblock \bibinfo{title}{The worldwide prevalence of adhd: is it an american
  condition?}
\newblock \bibinfo{journal}{World psychiatry} \bibinfo{volume}{2},
  \bibinfo{pages}{104}.
%Type = Article
\bibitem[{Forman and Zhang(2021)}]{forman2021targeting}
\bibinfo{author}{Forman, H.J.}, \bibinfo{author}{Zhang, H.},
  \bibinfo{year}{2021}.
\newblock \bibinfo{title}{Targeting oxidative stress in disease: Promise and
  limitations of antioxidant therapy}.
\newblock \bibinfo{journal}{Nature Reviews Drug Discovery}
  \bibinfo{volume}{20}, \bibinfo{pages}{689--709}.
%Type = Article
\bibitem[{Gellersen et~al.(2017)Gellersen, Guo, O?Callaghan, Tan, Sami and
  Hornberger}]{gellersen2017cerebellar}
\bibinfo{author}{Gellersen, H.M.}, \bibinfo{author}{Guo, C.C.},
  \bibinfo{author}{O?Callaghan, C.}, \bibinfo{author}{Tan, R.H.},
  \bibinfo{author}{Sami, S.}, \bibinfo{author}{Hornberger, M.},
  \bibinfo{year}{2017}.
\newblock \bibinfo{title}{Cerebellar atrophy in neurodegeneration?a
  meta-analysis}.
\newblock \bibinfo{journal}{Journal of Neurology, Neurosurgery \& Psychiatry}
  \bibinfo{volume}{88}, \bibinfo{pages}{780--788}.
%Type = Article
\bibitem[{Gonzalez-Ortiz et~al.(2023)Gonzalez-Ortiz, Turton, Kac, Smirnov,
  Premi, Ghidoni, Benussi, Cantoni, Saraceno, Rivolta
  et~al.}]{gonzalez2023brain}
\bibinfo{author}{Gonzalez-Ortiz, F.}, \bibinfo{author}{Turton, M.},
  \bibinfo{author}{Kac, P.R.}, \bibinfo{author}{Smirnov, D.},
  \bibinfo{author}{Premi, E.}, \bibinfo{author}{Ghidoni, R.},
  \bibinfo{author}{Benussi, L.}, \bibinfo{author}{Cantoni, V.},
  \bibinfo{author}{Saraceno, C.}, \bibinfo{author}{Rivolta, J.}, et~al.,
  \bibinfo{year}{2023}.
\newblock \bibinfo{title}{Brain-derived tau: a novel blood-based biomarker for
  alzheimer’s disease-type neurodegeneration}.
\newblock \bibinfo{journal}{Brain} \bibinfo{volume}{146},
  \bibinfo{pages}{1152--1165}.
%Type = Article
\bibitem[{Gu et~al.(2018)Gu, Wang, Kuen, Ma, Shahroudy, Shuai, Liu, Wang, Wang,
  Cai et~al.}]{gu2018recent}
\bibinfo{author}{Gu, J.}, \bibinfo{author}{Wang, Z.}, \bibinfo{author}{Kuen,
  J.}, \bibinfo{author}{Ma, L.}, \bibinfo{author}{Shahroudy, A.},
  \bibinfo{author}{Shuai, B.}, \bibinfo{author}{Liu, T.},
  \bibinfo{author}{Wang, X.}, \bibinfo{author}{Wang, G.}, \bibinfo{author}{Cai,
  J.}, et~al., \bibinfo{year}{2018}.
\newblock \bibinfo{title}{Recent advances in convolutional neural networks}.
\newblock \bibinfo{journal}{Pattern recognition} \bibinfo{volume}{77},
  \bibinfo{pages}{354--377}.
%Type = Article
\bibitem[{Hinshaw(2018)}]{hinshaw2018attention}
\bibinfo{author}{Hinshaw, S.P.}, \bibinfo{year}{2018}.
\newblock \bibinfo{title}{Attention deficit hyperactivity disorder (adhd):
  controversy, developmental mechanisms, and multiple levels of analysis}.
\newblock \bibinfo{journal}{Annual review of clinical psychology}
  \bibinfo{volume}{14}, \bibinfo{pages}{291--316}.
%Type = Article
\bibitem[{Huang et~al.(2020)Huang, Tan, Yang, Huang, Ou-Yang, Cao, Wang and
  Lei}]{huang2020}
\bibinfo{author}{Huang, F.}, \bibinfo{author}{Tan, E.L.},
  \bibinfo{author}{Yang, P.}, \bibinfo{author}{Huang, S.},
  \bibinfo{author}{Ou-Yang, L.}, \bibinfo{author}{Cao, J.},
  \bibinfo{author}{Wang, T.}, \bibinfo{author}{Lei, B.}, \bibinfo{year}{2020}.
\newblock \bibinfo{title}{Self-weighted adaptive structure learning for asd
  diagnosis via multi-template multi-center representation}.
\newblock \bibinfo{journal}{Medical Image Analysis} \bibinfo{volume}{63},
  \bibinfo{pages}{101662}.
%Type = Article
\bibitem[{Huang et~al.(2023)Huang, Wang, Ju, Shi, Ding and Zhang}]{huang2023sd}
\bibinfo{author}{Huang, J.}, \bibinfo{author}{Wang, M.}, \bibinfo{author}{Ju,
  H.}, \bibinfo{author}{Shi, Z.}, \bibinfo{author}{Ding, W.},
  \bibinfo{author}{Zhang, D.}, \bibinfo{year}{2023}.
\newblock \bibinfo{title}{Sd-cnn: A static-dynamic convolutional neural network
  for functional brain networks}.
\newblock \bibinfo{journal}{Medical Image Analysis} \bibinfo{volume}{83},
  \bibinfo{pages}{102679}.
%Type = Article
\bibitem[{Jacobs et~al.(2018)Jacobs, Hopkins, Mayrhofer, Bruner, van Leeuwen,
  Raaijmakers and Schmahmann}]{jacobs2018cerebellum}
\bibinfo{author}{Jacobs, H.I.}, \bibinfo{author}{Hopkins, D.A.},
  \bibinfo{author}{Mayrhofer, H.C.}, \bibinfo{author}{Bruner, E.},
  \bibinfo{author}{van Leeuwen, F.W.}, \bibinfo{author}{Raaijmakers, W.},
  \bibinfo{author}{Schmahmann, J.D.}, \bibinfo{year}{2018}.
\newblock \bibinfo{title}{The cerebellum in alzheimer?s disease: evaluating its
  role in cognitive decline}.
\newblock \bibinfo{journal}{Brain} \bibinfo{volume}{141},
  \bibinfo{pages}{37--47}.
%Type = Article
\bibitem[{Jie et~al.(2020)Jie, Liu, Lian, Shi and Shen}]{jie2020designing}
\bibinfo{author}{Jie, B.}, \bibinfo{author}{Liu, M.}, \bibinfo{author}{Lian,
  C.}, \bibinfo{author}{Shi, F.}, \bibinfo{author}{Shen, D.},
  \bibinfo{year}{2020}.
\newblock \bibinfo{title}{Designing weighted correlation kernels in
  convolutional neural networks for functional connectivity based brain disease
  diagnosis}.
\newblock \bibinfo{journal}{Medical image analysis} \bibinfo{volume}{63},
  \bibinfo{pages}{101709}.
%Type = Article
\bibitem[{Jie et~al.(2018)Jie, Liu and Shen}]{jie2018integration}
\bibinfo{author}{Jie, B.}, \bibinfo{author}{Liu, M.}, \bibinfo{author}{Shen,
  D.}, \bibinfo{year}{2018}.
\newblock \bibinfo{title}{Integration of temporal and spatial properties of
  dynamic connectivity networks for automatic diagnosis of brain disease}.
\newblock \bibinfo{journal}{Medical image analysis} \bibinfo{volume}{47},
  \bibinfo{pages}{81--94}.
%Type = Article
\bibitem[{Jie et~al.(2014)Jie, Zhang, Wee and Shen}]{jie2014topological}
\bibinfo{author}{Jie, B.}, \bibinfo{author}{Zhang, D.}, \bibinfo{author}{Wee,
  C.Y.}, \bibinfo{author}{Shen, D.}, \bibinfo{year}{2014}.
\newblock \bibinfo{title}{Topological graph kernel on multiple thresholded
  functional connectivity networks for mild cognitive impairment
  classification}.
\newblock \bibinfo{journal}{Human brain mapping} \bibinfo{volume}{35},
  \bibinfo{pages}{2876--2897}.
%Type = Article
\bibitem[{Kam et~al.(2019)Kam, Zhang, Jiao and Shen}]{kam2019deep}
\bibinfo{author}{Kam, T.E.}, \bibinfo{author}{Zhang, H.},
  \bibinfo{author}{Jiao, Z.}, \bibinfo{author}{Shen, D.}, \bibinfo{year}{2019}.
\newblock \bibinfo{title}{Deep learning of static and dynamic brain functional
  networks for early mci detection}.
\newblock \bibinfo{journal}{IEEE transactions on medical imaging}
  \bibinfo{volume}{39}, \bibinfo{pages}{478--487}.
%Type = Article
\bibitem[{Kamagata et~al.(2022)Kamagata, Andica, Takabayashi, Saito, Taoka,
  Nozaki, Kikuta, Fujita, Hagiwara, Kamiya et~al.}]{kamagata2022association}
\bibinfo{author}{Kamagata, K.}, \bibinfo{author}{Andica, C.},
  \bibinfo{author}{Takabayashi, K.}, \bibinfo{author}{Saito, Y.},
  \bibinfo{author}{Taoka, T.}, \bibinfo{author}{Nozaki, H.},
  \bibinfo{author}{Kikuta, J.}, \bibinfo{author}{Fujita, S.},
  \bibinfo{author}{Hagiwara, A.}, \bibinfo{author}{Kamiya, K.}, et~al.,
  \bibinfo{year}{2022}.
\newblock \bibinfo{title}{Association of mri indices of glymphatic system with
  amyloid deposition and cognition in mild cognitive impairment and alzheimer
  disease}.
\newblock \bibinfo{journal}{Neurology} \bibinfo{volume}{99},
  \bibinfo{pages}{e2648--e2660}.
%Type = Article
\bibitem[{Kavcic and Duffy(2003)}]{kavcic2003attentional}
\bibinfo{author}{Kavcic, V.}, \bibinfo{author}{Duffy, C.J.},
  \bibinfo{year}{2003}.
\newblock \bibinfo{title}{Attentional dynamics and visual perception:
  mechanisms of spatial disorientation in alzheimer’s disease}.
\newblock \bibinfo{journal}{Brain} \bibinfo{volume}{126},
  \bibinfo{pages}{1173--1181}.
%Type = Article
\bibitem[{Kawahara et~al.(2017)Kawahara, Brown, Miller, Booth, Chau, Grunau,
  Zwicker and Hamarneh}]{kawahara2017brainnetcnn}
\bibinfo{author}{Kawahara, J.}, \bibinfo{author}{Brown, C.J.},
  \bibinfo{author}{Miller, S.P.}, \bibinfo{author}{Booth, B.G.},
  \bibinfo{author}{Chau, V.}, \bibinfo{author}{Grunau, R.E.},
  \bibinfo{author}{Zwicker, J.G.}, \bibinfo{author}{Hamarneh, G.},
  \bibinfo{year}{2017}.
\newblock \bibinfo{title}{Brainnetcnn: Convolutional neural networks for brain
  networks; towards predicting neurodevelopment}.
\newblock \bibinfo{journal}{NeuroImage} \bibinfo{volume}{146},
  \bibinfo{pages}{1038--1049}.
%Type = Article
\bibitem[{Khojaste-Sarakhsi et~al.(2022)Khojaste-Sarakhsi, Haghighi, Ghomi and
  Marchiori}]{khojaste2022deep}
\bibinfo{author}{Khojaste-Sarakhsi, M.}, \bibinfo{author}{Haghighi, S.S.},
  \bibinfo{author}{Ghomi, S.F.}, \bibinfo{author}{Marchiori, E.},
  \bibinfo{year}{2022}.
\newblock \bibinfo{title}{Deep learning for alzheimer's disease diagnosis: A
  survey}.
\newblock \bibinfo{journal}{Artificial Intelligence in Medicine} ,
  \bibinfo{pages}{102332}.
%Type = Article
\bibitem[{Kshatri and Singh(2023)}]{kshatri2023convolutional}
\bibinfo{author}{Kshatri, S.S.}, \bibinfo{author}{Singh, D.},
  \bibinfo{year}{2023}.
\newblock \bibinfo{title}{Convolutional neural network in medical image
  analysis: a review}.
\newblock \bibinfo{journal}{Archives of Computational Methods in Engineering}
  \bibinfo{volume}{30}, \bibinfo{pages}{2793--2810}.
%Type = Article
\bibitem[{Kumfor et~al.(2013)Kumfor, Irish, Hodges and
  Piguet}]{kumfor2013orbitofrontal}
\bibinfo{author}{Kumfor, F.}, \bibinfo{author}{Irish, M.},
  \bibinfo{author}{Hodges, J.R.}, \bibinfo{author}{Piguet, O.},
  \bibinfo{year}{2013}.
\newblock \bibinfo{title}{The orbitofrontal cortex is involved in emotional
  enhancement of memory: evidence from the dementias}.
\newblock \bibinfo{journal}{Brain} \bibinfo{volume}{136},
  \bibinfo{pages}{2992--3003}.
%Type = Article
\bibitem[{Landin-Romero et~al.(2017)Landin-Romero, Kumfor, Leyton, Irish,
  Hodges and Piguet}]{landin2017disease}
\bibinfo{author}{Landin-Romero, R.}, \bibinfo{author}{Kumfor, F.},
  \bibinfo{author}{Leyton, C.E.}, \bibinfo{author}{Irish, M.},
  \bibinfo{author}{Hodges, J.R.}, \bibinfo{author}{Piguet, O.},
  \bibinfo{year}{2017}.
\newblock \bibinfo{title}{Disease-specific patterns of cortical and subcortical
  degeneration in a longitudinal study of alzheimer's disease and
  behavioural-variant frontotemporal dementia}.
\newblock \bibinfo{journal}{Neuroimage} \bibinfo{volume}{151},
  \bibinfo{pages}{72--80}.
%Type = Article
\bibitem[{Lei et~al.(2022)Lei, Zhang, Liu, Xu, Yue, Cao, Hu, Yu, Yang, Wang
  et~al.}]{lei2022longitudinal}
\bibinfo{author}{Lei, B.}, \bibinfo{author}{Zhang, Y.}, \bibinfo{author}{Liu,
  D.}, \bibinfo{author}{Xu, Y.}, \bibinfo{author}{Yue, G.},
  \bibinfo{author}{Cao, J.}, \bibinfo{author}{Hu, H.}, \bibinfo{author}{Yu,
  S.}, \bibinfo{author}{Yang, P.}, \bibinfo{author}{Wang, T.}, et~al.,
  \bibinfo{year}{2022}.
\newblock \bibinfo{title}{Longitudinal study of early mild cognitive impairment
  via similarity-constrained group learning and self-attention based sbi-lstm}.
\newblock \bibinfo{journal}{Knowledge-Based Systems} \bibinfo{volume}{254},
  \bibinfo{pages}{109466}.
%Type = Article
\bibitem[{Li et~al.(2021)Li, Pan and Carrasco}]{li2021different}
\bibinfo{author}{Li, H.H.}, \bibinfo{author}{Pan, J.},
  \bibinfo{author}{Carrasco, M.}, \bibinfo{year}{2021}.
\newblock \bibinfo{title}{Different computations underlie overt presaccadic and
  covert spatial attention}.
\newblock \bibinfo{journal}{Nature Human Behaviour} \bibinfo{volume}{5},
  \bibinfo{pages}{1418--1431}.
%Type = Article
\bibitem[{Liu et~al.(2022)Liu, Wang, Wang, Zhang and Xiong}]{liu2022enhanced}
\bibinfo{author}{Liu, L.}, \bibinfo{author}{Wang, Y.P.}, \bibinfo{author}{Wang,
  Y.}, \bibinfo{author}{Zhang, P.}, \bibinfo{author}{Xiong, S.},
  \bibinfo{year}{2022}.
\newblock \bibinfo{title}{An enhanced multi-modal brain graph network for
  classifying neuropsychiatric disorders}.
\newblock \bibinfo{journal}{Medical Image Analysis} \bibinfo{volume}{81},
  \bibinfo{pages}{102550}.
%Type = Article
\bibitem[{van Loenhoud et~al.(2019)van Loenhoud, van~der Flier, Wink, Dicks,
  Groot, Twisk, Barkhof, Scheltens, Ossenkoppele, Initiative
  et~al.}]{van2019cognitive}
\bibinfo{author}{van Loenhoud, A.C.}, \bibinfo{author}{van~der Flier, W.M.},
  \bibinfo{author}{Wink, A.M.}, \bibinfo{author}{Dicks, E.},
  \bibinfo{author}{Groot, C.}, \bibinfo{author}{Twisk, J.},
  \bibinfo{author}{Barkhof, F.}, \bibinfo{author}{Scheltens, P.},
  \bibinfo{author}{Ossenkoppele, R.}, \bibinfo{author}{Initiative, A.D.N.},
  et~al., \bibinfo{year}{2019}.
\newblock \bibinfo{title}{Cognitive reserve and clinical progression in
  alzheimer disease: a paradoxical relationship}.
\newblock \bibinfo{journal}{Neurology} \bibinfo{volume}{93},
  \bibinfo{pages}{e334--e346}.
%Type = Article
\bibitem[{Love and Miners(2016)}]{love2016cerebrovascular}
\bibinfo{author}{Love, S.}, \bibinfo{author}{Miners, J.S.},
  \bibinfo{year}{2016}.
\newblock \bibinfo{title}{Cerebrovascular disease in ageing and alzheimer’s
  disease}.
\newblock \bibinfo{journal}{Acta neuropathologica} \bibinfo{volume}{131},
  \bibinfo{pages}{645--658}.
%Type = Article
\bibitem[{Matthews and Hampshire(2016)}]{matthews2016clinical}
\bibinfo{author}{Matthews, P.M.}, \bibinfo{author}{Hampshire, A.},
  \bibinfo{year}{2016}.
\newblock \bibinfo{title}{Clinical concepts emerging from fmri functional
  connectomics}.
\newblock \bibinfo{journal}{Neuron} \bibinfo{volume}{91},
  \bibinfo{pages}{511--528}.
%Type = Article
\bibitem[{Matthews et~al.(2006)Matthews, Honey and
  Bullmore}]{matthews2006applications}
\bibinfo{author}{Matthews, P.M.}, \bibinfo{author}{Honey, G.D.},
  \bibinfo{author}{Bullmore, E.T.}, \bibinfo{year}{2006}.
\newblock \bibinfo{title}{Applications of fmri in translational medicine and
  clinical practice}.
\newblock \bibinfo{journal}{Nature Reviews Neuroscience} \bibinfo{volume}{7},
  \bibinfo{pages}{732--744}.
%Type = Article
\bibitem[{McDonough et~al.(2020)McDonough, Festini and
  Wood}]{mcdonough2020risk}
\bibinfo{author}{McDonough, I.M.}, \bibinfo{author}{Festini, S.B.},
  \bibinfo{author}{Wood, M.M.}, \bibinfo{year}{2020}.
\newblock \bibinfo{title}{Risk for alzheimer's disease: A review of long-term
  episodic memory encoding and retrieval fmri studies}.
\newblock \bibinfo{journal}{Ageing Research Reviews} \bibinfo{volume}{62},
  \bibinfo{pages}{101133}.
%Type = Article
\bibitem[{McDuff and Sumi(1985)}]{mcduff1985subcortical}
\bibinfo{author}{McDuff, T.}, \bibinfo{author}{Sumi, S.}, \bibinfo{year}{1985}.
\newblock \bibinfo{title}{Subcortical degeneration in alzheimer's disease}.
\newblock \bibinfo{journal}{Neurology} \bibinfo{volume}{35},
  \bibinfo{pages}{123--123}.
%Type = Article
\bibitem[{Merrill and Sabharwal(2023)}]{merrill2023parallelism}
\bibinfo{author}{Merrill, W.}, \bibinfo{author}{Sabharwal, A.},
  \bibinfo{year}{2023}.
\newblock \bibinfo{title}{The parallelism tradeoff: Limitations of
  log-precision transformers}.
\newblock \bibinfo{journal}{Transactions of the Association for Computational
  Linguistics} \bibinfo{volume}{11}, \bibinfo{pages}{531--545}.
%Type = Article
\bibitem[{Mishra and Verma(2022)}]{mishra2022graph}
\bibinfo{author}{Mishra, L.}, \bibinfo{author}{Verma, S.},
  \bibinfo{year}{2022}.
\newblock \bibinfo{title}{Graph attention autoencoder inspired cnn based brain
  tumor classification using mri}.
\newblock \bibinfo{journal}{Neurocomputing} \bibinfo{volume}{503},
  \bibinfo{pages}{236--247}.
%Type = Article
\bibitem[{Nandi et~al.(2022)Nandi, Counts, Chen, Seligman, Tortorice, Vigo and
  Bloom}]{nandi2022global}
\bibinfo{author}{Nandi, A.}, \bibinfo{author}{Counts, N.},
  \bibinfo{author}{Chen, S.}, \bibinfo{author}{Seligman, B.},
  \bibinfo{author}{Tortorice, D.}, \bibinfo{author}{Vigo, D.},
  \bibinfo{author}{Bloom, D.E.}, \bibinfo{year}{2022}.
\newblock \bibinfo{title}{Global and regional projections of the economic
  burden of alzheimer's disease and related dementias from 2019 to 2050: A
  value of statistical life approach}.
\newblock \bibinfo{journal}{EClinicalMedicine} \bibinfo{volume}{51}.
%Type = Article
\bibitem[{Noble(2006)}]{noble2006support}
\bibinfo{author}{Noble, W.S.}, \bibinfo{year}{2006}.
\newblock \bibinfo{title}{What is a support vector machine?}
\newblock \bibinfo{journal}{Nature biotechnology} \bibinfo{volume}{24},
  \bibinfo{pages}{1565--1567}.
%Type = Article
\bibitem[{Pan et~al.(2023)Pan, Zeng, Yang, Lai, Hu, Song, Jiang and
  (ADNI)}]{pan2023deep}
\bibinfo{author}{Pan, D.}, \bibinfo{author}{Zeng, A.}, \bibinfo{author}{Yang,
  B.}, \bibinfo{author}{Lai, G.}, \bibinfo{author}{Hu, B.},
  \bibinfo{author}{Song, X.}, \bibinfo{author}{Jiang, T.},
  \bibinfo{author}{(ADNI), A.D.N.I.}, \bibinfo{year}{2023}.
\newblock \bibinfo{title}{Deep learning for brain mri confirms patterned
  pathological progression in alzheimer's disease}.
\newblock \bibinfo{journal}{Advanced Science} \bibinfo{volume}{10},
  \bibinfo{pages}{2204717}.
%Type = Inproceedings
\bibitem[{Paul and Chen(2022)}]{paul2022vision}
\bibinfo{author}{Paul, S.}, \bibinfo{author}{Chen, P.Y.}, \bibinfo{year}{2022}.
\newblock \bibinfo{title}{Vision transformers are robust learners}, in:
  \bibinfo{booktitle}{Proceedings of the AAAI conference on Artificial
  Intelligence}, pp. \bibinfo{pages}{2071--2081}.
%Type = Article
\bibitem[{Penniello et~al.(1995)Penniello, Lambert, Eustache, Petit-Tabou{\'e},
  Barr{\'e}, Viader, Morin, Lechevalier and Baron}]{penniello1995pet}
\bibinfo{author}{Penniello, M.J.}, \bibinfo{author}{Lambert, J.},
  \bibinfo{author}{Eustache, F.}, \bibinfo{author}{Petit-Tabou{\'e}, M.C.},
  \bibinfo{author}{Barr{\'e}, L.}, \bibinfo{author}{Viader, F.},
  \bibinfo{author}{Morin, P.}, \bibinfo{author}{Lechevalier, B.},
  \bibinfo{author}{Baron, J.C.}, \bibinfo{year}{1995}.
\newblock \bibinfo{title}{A pet study of the functional neuroanatomy of writing
  impairment in alzheimer's disease the role of the left supramarginal and left
  angular gyri}.
\newblock \bibinfo{journal}{Brain} \bibinfo{volume}{118},
  \bibinfo{pages}{697--706}.
%Type = Article
\bibitem[{Perovnik et~al.(2023)Perovnik, Rus, Schindlbeck and
  Eidelberg}]{perovnik2023functional}
\bibinfo{author}{Perovnik, M.}, \bibinfo{author}{Rus, T.},
  \bibinfo{author}{Schindlbeck, K.A.}, \bibinfo{author}{Eidelberg, D.},
  \bibinfo{year}{2023}.
\newblock \bibinfo{title}{Functional brain networks in the evaluation of
  patients with neurodegenerative disorders}.
\newblock \bibinfo{journal}{Nature Reviews Neurology} \bibinfo{volume}{19},
  \bibinfo{pages}{73--90}.
%Type = Article
\bibitem[{Petersen and Posner(2012)}]{petersen2012attention}
\bibinfo{author}{Petersen, S.E.}, \bibinfo{author}{Posner, M.I.},
  \bibinfo{year}{2012}.
\newblock \bibinfo{title}{The attention system of the human brain: 20 years
  after}.
\newblock \bibinfo{journal}{Annual review of neuroscience}
  \bibinfo{volume}{35}, \bibinfo{pages}{73--89}.
%Type = Article
\bibitem[{Phan et~al.(2023)Phan, Nguyen and Hwang}]{phan2023aspect}
\bibinfo{author}{Phan, H.T.}, \bibinfo{author}{Nguyen, N.T.},
  \bibinfo{author}{Hwang, D.}, \bibinfo{year}{2023}.
\newblock \bibinfo{title}{Aspect-level sentiment analysis: A survey of graph
  convolutional network methods}.
\newblock \bibinfo{journal}{Information Fusion} \bibinfo{volume}{91},
  \bibinfo{pages}{149--172}.
%Type = Article
\bibitem[{Posner and Petersen(1990)}]{posner1990attention}
\bibinfo{author}{Posner, M.I.}, \bibinfo{author}{Petersen, S.E.},
  \bibinfo{year}{1990}.
\newblock \bibinfo{title}{The attention system of the human brain}.
\newblock \bibinfo{journal}{Annual review of neuroscience}
  \bibinfo{volume}{13}, \bibinfo{pages}{25--42}.
%Type = Article
\bibitem[{Qiu et~al.(2022)Qiu, Miller, Joshi, Lee, Xue, Ni, Wang,
  De~Anda-Duran, Hwang, Cramer et~al.}]{qiu2022multimodal}
\bibinfo{author}{Qiu, S.}, \bibinfo{author}{Miller, M.I.},
  \bibinfo{author}{Joshi, P.S.}, \bibinfo{author}{Lee, J.C.},
  \bibinfo{author}{Xue, C.}, \bibinfo{author}{Ni, Y.}, \bibinfo{author}{Wang,
  Y.}, \bibinfo{author}{De~Anda-Duran, I.}, \bibinfo{author}{Hwang, P.H.},
  \bibinfo{author}{Cramer, J.A.}, et~al., \bibinfo{year}{2022}.
\newblock \bibinfo{title}{Multimodal deep learning for alzheimer’s disease
  dementia assessment}.
\newblock \bibinfo{journal}{Nature communications} \bibinfo{volume}{13},
  \bibinfo{pages}{3404}.
%Type = Article
\bibitem[{Rabinovich et~al.(2012)Rabinovich, Afraimovich, Bick and
  Varona}]{rabinovich2012information}
\bibinfo{author}{Rabinovich, M.I.}, \bibinfo{author}{Afraimovich, V.S.},
  \bibinfo{author}{Bick, C.}, \bibinfo{author}{Varona, P.},
  \bibinfo{year}{2012}.
\newblock \bibinfo{title}{Information flow dynamics in the brain}.
\newblock \bibinfo{journal}{Physics of life reviews} \bibinfo{volume}{9},
  \bibinfo{pages}{51--73}.
%Type = Article
\bibitem[{Scheltens et~al.(2021)Scheltens, De~Strooper, Kivipelto, Holstege,
  Ch{\'e}telat, Teunissen, Cummings and van~der Flier}]{scheltens2021alzheimer}
\bibinfo{author}{Scheltens, P.}, \bibinfo{author}{De~Strooper, B.},
  \bibinfo{author}{Kivipelto, M.}, \bibinfo{author}{Holstege, H.},
  \bibinfo{author}{Ch{\'e}telat, G.}, \bibinfo{author}{Teunissen, C.E.},
  \bibinfo{author}{Cummings, J.}, \bibinfo{author}{van~der Flier, W.M.},
  \bibinfo{year}{2021}.
\newblock \bibinfo{title}{Alzheimer's disease}.
\newblock \bibinfo{journal}{The Lancet} \bibinfo{volume}{397},
  \bibinfo{pages}{1577--1590}.
%Type = Article
\bibitem[{Schmahmann(2016)}]{schmahmann2016cerebellum}
\bibinfo{author}{Schmahmann, J.D.}, \bibinfo{year}{2016}.
\newblock \bibinfo{title}{Cerebellum in alzheimer’s disease and
  frontotemporal dementia: not a silent bystander}.
\newblock \bibinfo{journal}{Brain} \bibinfo{volume}{139},
  \bibinfo{pages}{1314--1318}.
%Type = Article
\bibitem[{Shireby et~al.(2022)Shireby, Dempster, Policicchio, Smith, Pishva,
  Chioza, Davies, Burrage, Lunnon, Seiler~Vellame et~al.}]{shireby2022dna}
\bibinfo{author}{Shireby, G.}, \bibinfo{author}{Dempster, E.L.},
  \bibinfo{author}{Policicchio, S.}, \bibinfo{author}{Smith, R.G.},
  \bibinfo{author}{Pishva, E.}, \bibinfo{author}{Chioza, B.},
  \bibinfo{author}{Davies, J.P.}, \bibinfo{author}{Burrage, J.},
  \bibinfo{author}{Lunnon, K.}, \bibinfo{author}{Seiler~Vellame, D.}, et~al.,
  \bibinfo{year}{2022}.
\newblock \bibinfo{title}{Dna methylation signatures of alzheimer’s disease
  neuropathology in the cortex are primarily driven by variation in
  non-neuronal cell-types}.
\newblock \bibinfo{journal}{Nature Communications} \bibinfo{volume}{13},
  \bibinfo{pages}{5620}.
%Type = Article
\bibitem[{Shoeibi et~al.(2022)Shoeibi, Khodatars, Jafari, Ghassemi, Moridian,
  Alizadesani, Ling, Khosravi, Alinejad-Rokny, Lam
  et~al.}]{shoeibi2022diagnosis}
\bibinfo{author}{Shoeibi, A.}, \bibinfo{author}{Khodatars, M.},
  \bibinfo{author}{Jafari, M.}, \bibinfo{author}{Ghassemi, N.},
  \bibinfo{author}{Moridian, P.}, \bibinfo{author}{Alizadesani, R.},
  \bibinfo{author}{Ling, S.H.}, \bibinfo{author}{Khosravi, A.},
  \bibinfo{author}{Alinejad-Rokny, H.}, \bibinfo{author}{Lam, H.}, et~al.,
  \bibinfo{year}{2022}.
\newblock \bibinfo{title}{Diagnosis of brain diseases in fusion of neuroimaging
  modalities using deep learning: A review}.
\newblock \bibinfo{journal}{Information Fusion} .
%Type = Article
\bibitem[{Sri and Mallikarjunarao(2022)}]{sri2022detection}
\bibinfo{author}{Sri, M.S.C.}, \bibinfo{author}{Mallikarjunarao, P.},
  \bibinfo{year}{2022}.
\newblock \bibinfo{title}{Detection of alzheimer disease usingdeep learning
  cnn}.
\newblock \bibinfo{journal}{blood} \bibinfo{volume}{6}, \bibinfo{pages}{8}.
%Type = Inproceedings
\bibitem[{Szegedy et~al.(2017)Szegedy, Ioffe, Vanhoucke and
  Alemi}]{szegedy2017inception}
\bibinfo{author}{Szegedy, C.}, \bibinfo{author}{Ioffe, S.},
  \bibinfo{author}{Vanhoucke, V.}, \bibinfo{author}{Alemi, A.A.},
  \bibinfo{year}{2017}.
\newblock \bibinfo{title}{Inception-v4, inception-resnet and the impact of
  residual connections on learning}.
%Type = Article
\bibitem[{Toniolo et~al.(2018)Toniolo, Serra, Olivito, Marra, Bozzali and
  Cercignani}]{toniolo2018patterns}
\bibinfo{author}{Toniolo, S.}, \bibinfo{author}{Serra, L.},
  \bibinfo{author}{Olivito, G.}, \bibinfo{author}{Marra, C.},
  \bibinfo{author}{Bozzali, M.}, \bibinfo{author}{Cercignani, M.},
  \bibinfo{year}{2018}.
\newblock \bibinfo{title}{Patterns of cerebellar gray matter atrophy across
  alzheimer?s disease progression}.
\newblock \bibinfo{journal}{Frontiers in Cellular Neuroscience}
  \bibinfo{volume}{12}, \bibinfo{pages}{430}.
%Type = Article
\bibitem[{Uddin et~al.(2013)Uddin, Supekar, Lynch, Khouzam, Phillips,
  Feinstein, Ryali and Menon}]{uddin2013salience}
\bibinfo{author}{Uddin, L.Q.}, \bibinfo{author}{Supekar, K.},
  \bibinfo{author}{Lynch, C.J.}, \bibinfo{author}{Khouzam, A.},
  \bibinfo{author}{Phillips, J.}, \bibinfo{author}{Feinstein, C.},
  \bibinfo{author}{Ryali, S.}, \bibinfo{author}{Menon, V.},
  \bibinfo{year}{2013}.
\newblock \bibinfo{title}{Salience network--based classification and prediction
  of symptom severity in children with autism}.
\newblock \bibinfo{journal}{JAMA psychiatry} \bibinfo{volume}{70},
  \bibinfo{pages}{869--879}.
%Type = Inproceedings
\bibitem[{Wang and Liu(2021)}]{wang2021understanding}
\bibinfo{author}{Wang, F.}, \bibinfo{author}{Liu, H.}, \bibinfo{year}{2021}.
\newblock \bibinfo{title}{Understanding the behaviour of contrastive loss}, in:
  \bibinfo{booktitle}{Proceedings of the IEEE/CVF conference on computer vision
  and pattern recognition}, pp. \bibinfo{pages}{2495--2504}.
%Type = Article
\bibitem[{Wee et~al.(2011)Wee, Yap, Li, Denny, Browndyke, Potter, Welsh-Bohmer,
  Wang and Shen}]{wee2011enriched}
\bibinfo{author}{Wee, C.Y.}, \bibinfo{author}{Yap, P.T.}, \bibinfo{author}{Li,
  W.}, \bibinfo{author}{Denny, K.}, \bibinfo{author}{Browndyke, J.N.},
  \bibinfo{author}{Potter, G.G.}, \bibinfo{author}{Welsh-Bohmer, K.A.},
  \bibinfo{author}{Wang, L.}, \bibinfo{author}{Shen, D.}, \bibinfo{year}{2011}.
\newblock \bibinfo{title}{Enriched white matter connectivity networks for
  accurate identification of mci patients}.
\newblock \bibinfo{journal}{Neuroimage} \bibinfo{volume}{54},
  \bibinfo{pages}{1812--1822}.
%Type = Article
\bibitem[{Wright et~al.(2007)Wright, Dickerson, Feczko, Negeira and
  Williams}]{wright2007functional}
\bibinfo{author}{Wright, C.I.}, \bibinfo{author}{Dickerson, B.C.},
  \bibinfo{author}{Feczko, E.}, \bibinfo{author}{Negeira, A.},
  \bibinfo{author}{Williams, D.}, \bibinfo{year}{2007}.
\newblock \bibinfo{title}{A functional magnetic resonance imaging study of
  amygdala responses to human faces in aging and mild alzheimer’s disease}.
\newblock \bibinfo{journal}{Biological psychiatry} \bibinfo{volume}{62},
  \bibinfo{pages}{1388--1395}.
%Type = Article
\bibitem[{Yan et~al.(2019)Yan, Calhoun, Song, Cui, Yan, Liu, Fan, Zuo, Yang, Xu
  et~al.}]{yan2019discriminating}
\bibinfo{author}{Yan, W.}, \bibinfo{author}{Calhoun, V.},
  \bibinfo{author}{Song, M.}, \bibinfo{author}{Cui, Y.}, \bibinfo{author}{Yan,
  H.}, \bibinfo{author}{Liu, S.}, \bibinfo{author}{Fan, L.},
  \bibinfo{author}{Zuo, N.}, \bibinfo{author}{Yang, Z.}, \bibinfo{author}{Xu,
  K.}, et~al., \bibinfo{year}{2019}.
\newblock \bibinfo{title}{Discriminating schizophrenia using recurrent neural
  network applied on time courses of multi-site fmri data}.
\newblock \bibinfo{journal}{EBioMedicine} \bibinfo{volume}{47},
  \bibinfo{pages}{543--552}.
%Type = Article
\bibitem[{Yang et~al.(2019)Yang, Zhou, Ni, Xu, Chen, Wang and
  Lei}]{yang2019fused}
\bibinfo{author}{Yang, P.}, \bibinfo{author}{Zhou, F.}, \bibinfo{author}{Ni,
  D.}, \bibinfo{author}{Xu, Y.}, \bibinfo{author}{Chen, S.},
  \bibinfo{author}{Wang, T.}, \bibinfo{author}{Lei, B.}, \bibinfo{year}{2019}.
\newblock \bibinfo{title}{Fused sparse network learning for longitudinal
  analysis of mild cognitive impairment}.
\newblock \bibinfo{journal}{IEEE transactions on cybernetics}
  \bibinfo{volume}{51}, \bibinfo{pages}{233--246}.
%Type = Article
\bibitem[{Zhang et~al.(2022a)Zhang, Song, Wang, Zhang, Wang, Wang and
  Zhang}]{zhang2022classification}
\bibinfo{author}{Zhang, H.}, \bibinfo{author}{Song, R.}, \bibinfo{author}{Wang,
  L.}, \bibinfo{author}{Zhang, L.}, \bibinfo{author}{Wang, D.},
  \bibinfo{author}{Wang, C.}, \bibinfo{author}{Zhang, W.},
  \bibinfo{year}{2022}a.
\newblock \bibinfo{title}{Classification of brain disorders in rs-fmri via
  local-to-global graph neural networks}.
\newblock \bibinfo{journal}{IEEE Transactions on Medical Imaging}
  \bibinfo{volume}{42}, \bibinfo{pages}{444--455}.
%Type = Article
\bibitem[{Zhang et~al.(2023)Zhang, Li, Yin, Zhang and
  Grzegorzek}]{zhang2023applications}
\bibinfo{author}{Zhang, J.}, \bibinfo{author}{Li, C.}, \bibinfo{author}{Yin,
  Y.}, \bibinfo{author}{Zhang, J.}, \bibinfo{author}{Grzegorzek, M.},
  \bibinfo{year}{2023}.
\newblock \bibinfo{title}{Applications of artificial neural networks in
  microorganism image analysis: a comprehensive review from conventional
  multilayer perceptron to popular convolutional neural network and potential
  visual transformer}.
\newblock \bibinfo{journal}{Artificial Intelligence Review}
  \bibinfo{volume}{56}, \bibinfo{pages}{1013--1070}.
%Type = Article
\bibitem[{Zhang et~al.(2022b)Zhang, Zhou, Wang, Liu and
  Shen}]{zhang2022diffusion}
\bibinfo{author}{Zhang, J.}, \bibinfo{author}{Zhou, L.}, \bibinfo{author}{Wang,
  L.}, \bibinfo{author}{Liu, M.}, \bibinfo{author}{Shen, D.},
  \bibinfo{year}{2022}b.
\newblock \bibinfo{title}{Diffusion kernel attention network for brain disorder
  classification}.
\newblock \bibinfo{journal}{IEEE Transactions on Medical Imaging}
  \bibinfo{volume}{41}, \bibinfo{pages}{2814--2827}.
%Type = Article
\bibitem[{Zhao et~al.(2022a)Zhao, Zhang, Thung, Mao, Lee and
  Shen}]{zhao2021diagnosis}
\bibinfo{author}{Zhao, F.}, \bibinfo{author}{Zhang, X.},
  \bibinfo{author}{Thung, K.H.}, \bibinfo{author}{Mao, N.},
  \bibinfo{author}{Lee, S.W.}, \bibinfo{author}{Shen, D.},
  \bibinfo{year}{2022}a.
\newblock \bibinfo{title}{Constructing multi-view high-order functional
  connectivity networks for diagnosis of autism spectrum disorder}.
\newblock \bibinfo{journal}{IEEE Transactions on Biomedical Engineering}
  \bibinfo{volume}{69}, \bibinfo{pages}{1237--1250}.
%Type = Article
\bibitem[{Zhao et~al.(2022b)Zhao, Yan, Luo, Zhi, Fu, Du, Yu, Jiang, Calhoun and
  Sui}]{zhao2022attention}
\bibinfo{author}{Zhao, M.}, \bibinfo{author}{Yan, W.}, \bibinfo{author}{Luo,
  N.}, \bibinfo{author}{Zhi, D.}, \bibinfo{author}{Fu, Z.},
  \bibinfo{author}{Du, Y.}, \bibinfo{author}{Yu, S.}, \bibinfo{author}{Jiang,
  T.}, \bibinfo{author}{Calhoun, V.D.}, \bibinfo{author}{Sui, J.},
  \bibinfo{year}{2022}b.
\newblock \bibinfo{title}{An attention-based hybrid deep learning framework
  integrating brain connectivity and activity of resting-state functional mri
  data}.
\newblock \bibinfo{journal}{Medical image analysis} \bibinfo{volume}{78},
  \bibinfo{pages}{102413}.
%Type = Article
\bibitem[{Zhu et~al.(2022)Zhu, Ma, Yuan and Zhu}]{zhu2022interpretable}
\bibinfo{author}{Zhu, Y.}, \bibinfo{author}{Ma, J.}, \bibinfo{author}{Yuan,
  C.}, \bibinfo{author}{Zhu, X.}, \bibinfo{year}{2022}.
\newblock \bibinfo{title}{Interpretable learning based dynamic graph
  convolutional networks for alzheimer’s disease analysis}.
\newblock \bibinfo{journal}{Information Fusion} \bibinfo{volume}{77},
  \bibinfo{pages}{53--61}.

\end{thebibliography}

\end{document}